\documentclass[11pt,a4paper]{article}
\pdfoutput=1

\usepackage{jheppub}

\usepackage{amsfonts}%
\usepackage{amsmath}
\usepackage{amssymb}
\usepackage[english]{babel}
\usepackage{bbold}
\usepackage{bm}
\usepackage{braket}
\usepackage{hhline}
\usepackage{color,graphicx}
\usepackage{hyperref}
\usepackage{ifpdf}
\usepackage[utf8]{inputenc}
\usepackage{slashed}
\usepackage{url}

\usepackage{tikz-feynman}

\definecolor{blue}{rgb}{0.1,0.1,0.8}
\definecolor{nicered}{rgb}{0.7,0.1,0.1}
\definecolor{nicegreen}{rgb}{0.1,0.5,0.1}
\hypersetup{colorlinks,citecolor= blue,linkcolor= blue}

\newcommand {\e}[1]{\mathrm{~#1}}

\newcommand{\Br}[0]{\mathrm{Br}}

\newcommand{\mc}[1]{\mathcal{#1}}

\newcommand{\beq}{\begin{equation}}
\newcommand{\eeq}{\end{equation}}
\newcommand{\bea}{\begin{eqnarray}}
\newcommand{\eea}{\end{eqnarray}}

\definecolor{Red}{rgb}{1.,0.,0.}

\newcommand{\Blue}[1]{{\color{blue}{#1}}}

\newcommand{\mrm}[1]{\mathrm{#1}}

\newcommand{\lagr}{\mathcal{L}}
\newcommand{\amp}{\mathcal{M}}

\newcommand{\lp}{\bm{p}_{\ell^+}}
\newcommand{\lm}{\bm{p}_{\ell^-}}
\renewcommand{\b}{\bm{p}_b}
\newcommand{\bbar}{\bm{p}_{\bar{b}}}
\newcommand{\h}{\bm{p}_h}

\newcommand{\hvec}[1]{\bm{\hat #1}}

\begin{document}

\arxivnumber{1909.00007}

\author[a]{Darius A. Faroughy}
\author[a,b]{Jernej F.\ Kamenik} 
\author[a,b]{Nejc Ko\v snik}
\author[a]{Aleks Smolkovi\v c}

\emailAdd{darius.faroughy@ijs.si}
\emailAdd{jernej.kamenik@ijs.si}
\emailAdd{nejc.kosnik@ijs.si}
\emailAdd{aleks.smolkovic@ijs.si}

\affiliation[a]{J. Stefan Institute, Jamova 39, P. O. Box 3000, 1001
  Ljubljana, Slovenia}
\affiliation[b]{Department of Physics,
  University of Ljubljana, Jadranska 19, 1000 Ljubljana, Slovenia}

\title{\Blue{Probing the $CP$ nature of the top quark Yukawa at hadron colliders}}

\date{\today}

\abstract{ We analyze the prospects of probing the $CP$-odd
  $i \tilde \kappa \bar t \gamma^5 t h$ interaction at the LHC and its
  projected upgrades, the high-luminosity and high-energy LHC,
  directly using associated on-shell Higgs boson and top quark or top
  quark pair production.  To this end we first construct a $CP$-odd
  observable based on top quark polarization in $W b \to t h$
  scattering with optimal linear sensitivity to $\tilde \kappa$.  For
  the corresponding hadronic process $pp \to t h j$ we present a
  method of extracting the phase-space dependent weight function that
  allows to retain close to optimal sensitivity to
  $\tilde{\kappa}$. We project future sensitivity to the signal in
  $pp \to t(\to \ell \nu b )h(\to b\bar b) j$. We also propose novel
  $CP$-odd observables for top quark pair production in association
  with the Higgs, $pp \to t \bar t h$, with semileptonically decaying
  tops and $h\to b \bar b$, that rely solely on measuring the momenta
  of leptons and $b$-jets from the decaying tops without having to
  distinguish the charge of the $b$-jets. Among the many possibilities
  we single out an observable that can potentially probe
  $\tilde \kappa \sim 0.5$ at the high-luminosity LHC and
  $\tilde \kappa \sim 0.1$ at high-energy LHC with $2\sigma$ confidence.}

\maketitle

%
\section{Introduction}
%
The coupling of the 125 GeV Higgs boson ($h$) to the top quark, which is the largest of the Standard Model (SM) couplings, is an important target for the LHC experiments. $CP$-violating $h$ couplings are particularly interesting as any sign of $CP$ violation in Higgs processes would constitute an indisputable New Physics (NP) signal. Existing data on Higgs production and decays is already precise enough to constrain any isolated modification of the top Yukawa to $\mathcal O(1)$~\cite{Ellis:2013yxa,Khachatryan:2016vau,Bhattacharyya:2012tj}. However, all existing measurements are based on $CP$-even observables with very limited sensitivity to $CP$-odd modifications of the top quark Yukawa. In principle, indirect collider bounds from Higgs decay and production ($gg \to h$, $h \to \gamma\gamma$), and especially the low-energy bounds on electric dipole moments (EDMs) of atoms and nuclei that target specifically $CP$-odd effects~\cite{Brod:2013cka,Ellis:2013yxa,Boudjema:2015nda}, are currently more constraining than direct collider probes. However, these constraints are subject to assumptions about other Higgs interactions, and in particular in the case of EDMs also other contributions unrelated to the Higgs.

A few existing proposals for LHC measurements of top quark pair production in association with the Higgs boson have studied manifestly $CP$-odd observables with on-shell $t$, $\bar t$ and $h$~\cite{hep-ph/9501339, Ellis:2013yxa, 1407.5089, Boudjema:2015nda, 1507.07926, 1603.03632, Gritsan:2016hjl, Li:2017dyz, AmorDosSantos:2017ayi, 1804.05874} (for similar studies at $e^+e^-$ colliders see e.g. Refs.~\cite{hep-ph/9605326,BhupalDev:2007ftb}). In particular, the $CP$ nature of the top-Higgs coupling in this case is reflected in the correlation between the spins of the tops, which can be reconstructed using the angular distributions of the top quark decay products. It turns out however that the resulting effects are typically almost prohibitively difficult to measure at the LHC due to limitations of simultaneous top quarks' reconstruction, as well as their spin and charge identification. An alternative is offered by the single top production with associated Higgs based on the hard process $b W \to t h$ and observed as $pp \to t H j$. Owing to simpler kinematics, the top quark polarization is more directly accessible in this case. In particular it can be reconstructed in semileptonic top decays through the angular distribution of the charged lepton in the top rest frame. Several existing studies in this direction have already proposed top quark polarization related observables~\cite{Ellis:2013yxa,Kobakhidze:2014gqa,1410.2701,1504.00611,1807.00281,Kraus:2019myc} in single top-Higgs associated production (see Ref.~\cite{Coleppa:2017rgb} for a similar analysis at a $pe$ collider). Yet while the literature abounds with proposals of $CP$-sensitive measurements both in $t\bar{t} h$ and $t h$ channels, there has been no study to systematically search for and construct observables with optimal sensitivity to the CP-odd top Yukawa under realistic conditions at hadron colliders.

In this paper we address this challenge by identifying observables with optimal sensitivity to a single $CP$-odd parameter in both $th$ and $t \bar{t} h$ associated production at the LHC, which can be realistically measured and exhibit close to optimal sensitivity to $CP$-odd interactions between the Higgs boson and the top quark. The proposed observables in $th$ are based on optimization of top-spin correlations previously studied in $t\bar t$ production~\cite{Fajfer:2012si}. In the case of $\bar t t h$ this procedure becomes intractable in practice and our construction relies instead on $CP$- and $P$-symmetry arguments.

To set the stage we write the effective top quark -- Higgs boson interaction as
\begin{equation}
\label{eq:YukawaTopHiggs}
\begin{split}   
\lagr_{ht} &= -\frac{y_t}{\sqrt{2}} \bar t(\kappa  + i \tilde{\kappa} \gamma_5) t h\,,
\end{split}
\end{equation} 
where $y_t = \sqrt{2} m_t/v$ is the top quark Yukawa in the SM, while
real dimensionless quantities $\kappa$, $\tilde \kappa$ parametrize
departures from the SM (at $\kappa =1$, $\tilde{\kappa} =0$). In the
context of the SM Lagrangian complemented by dimension-6 effective
interactions, $\tilde \kappa$ is generated from the operator
$|H|^2 \bar Q \tilde H u_R$ which decouples the Higgs couplings~\eqref{eq:YukawaTopHiggs} from
the quark mass matrix~\cite{AguilarSaavedra:2009mx}. Clearly, any
indication of a non-vanishing $\tilde{\kappa}$ would be an
indisputable sign of NP.  Our goal is to construct optimized and
practically measurable observables which probe the $CP$-odd parameter
$\tilde{\kappa}$ directly.

The rest of the paper is structured as follows. In Sec.~\ref{sec:thj} we study optimized top spin observables in single top quark and Higgs boson associated production, both in idealized partonic $Wb \to th$ scattering, which is tractable analytically, as well as in more realistic simulations of $thj$ production and reconstruction in proton collisions at $pp$ colliders. Sec.~\ref{sec:tth} contains the analysis of $CP$ violating observables built from accessible momenta in $t\bar t h$ production, both at partonic Monte Carlo~(MC) level and after including the background, reconstruction, and detector effects. Finally, our main conclusions are summarized in Sec.~\ref{sec:conclusions}.

%
\section{Optimized spin observable in $pp \to t h j$}\label{sec:thj}
%

\subsection{Parton level $Wb \to th$ analysis}
\begin{figure}
	\centering\includegraphics[scale=0.6]{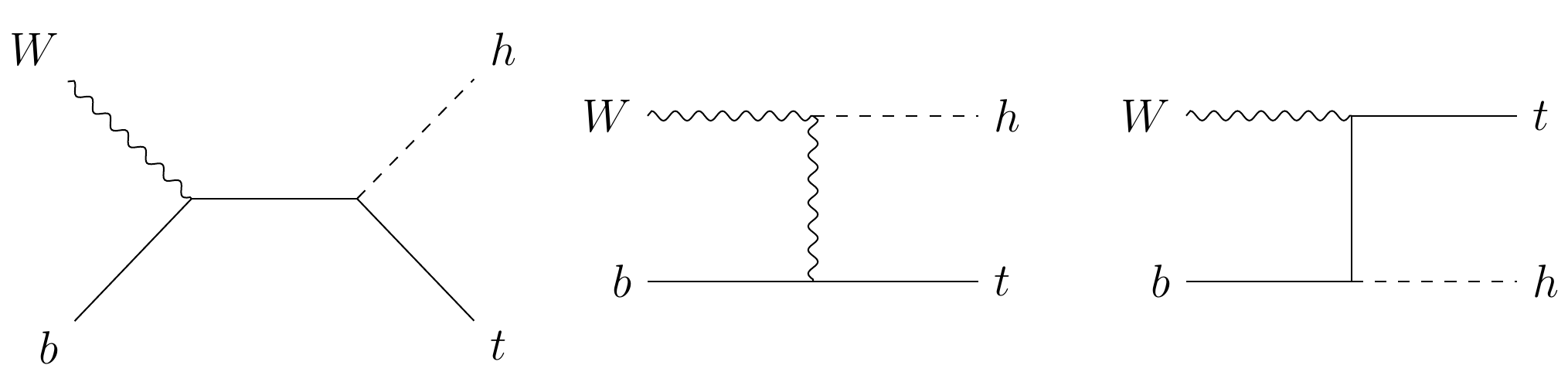}
	\caption{Tree-level diagrams contributing to $W  b  \to t  h$.}
	\label{fig:diagramsWbth}
\end{figure}

We begin by studying the effects of $\tilde \kappa$ on top spin
observables in the idealized case of $W (p) b (p') \to t (k) h (k')$
scattering, where the complete polarized scattering amplitudes can be
found in a compact analytic form.  This process can actually be
connected to a more realistic $pp \to t h j$ production in the high
energy limit, where the $W$ and $b$ quark mass effects are negligible
and the collinear emission of both initial state `partons' can be
described by the corresponding parton distribution functions.\footnote{See e.g. Sec.~3 of Ref.~\cite{Mangano:2016jyj} for an extended discussion on the validity of this approximation.}  Three
diagrams contribute to such parton level Higgs-top production in the
SM, shown in Fig.~\ref{fig:diagramsWbth}. Neglecting furthermore the
mass (and thus the corresponding Yukawa coupling) of the bottom quark,
we consider only the first two of the diagrams in
Fig.~\ref{fig:diagramsWbth}. The formalism presented here is based on
Refs.~\cite{Dicus:1984fu,hep-ph/0403035,Fajfer:2012si}. First, we
introduce the spin projection operator
\begin{equation}
\label{eq:spinprojector}
P(s) = \frac{1}{2} (1 + \gamma_5 \slashed{s}),
\end{equation}
where $s^\mu$ is a top spin four-vector, defined in a general frame as
\begin{equation}
\label{eq:spinFV}
s^\mu = \left(\frac{\bm{k}\cdot \hvec{s}}{m_t}, \hvec{s} + \frac{\bm{k} (\bm{k} \cdot \hvec{s})}{m_t (E_t + m_t)}\right).
\end{equation}
Vector $\bm{k}$ is the top quark momentum and $\hvec{s}$ is an
arbitrary unit vector. The physical significance of $\hvec{s}$ is
revealed if we make a rotation-free boost\footnote{The spatial component of four-vector $x^\mu$ transforms as $\bm{x}^* = \bm{x} + \left(\frac{\bm{x}\cdot \bm{k}}{m_t (E_t+m_t)}-\frac{x^0}{m_t} \right) \bm{k}$ upon a boost to the top rest frame.} to the top rest
frame where we find $s^{*\mu} = (0,\hvec{s})$. Therefore $s^2 = -1$,
$s \cdot k =0$, and $\hvec{s}$ corresponds to the polarization of the top quark in its rest frame. Projection onto a well
defined polarization of the top quark is achieved by inserting the
operator~\eqref{eq:spinprojector} at the amplitude level and
leads to the following relation at the cross-section level:
\begin{equation}
\label{eq:projections}
\begin{split}
u(p,s) \bar{u} (p,s) &= (\slashed{p}+m) P(+s), \\
v(p,s) \bar{v} (p,s) &= (\slashed{p}-m) P(-s).
\end{split}
\end{equation}
Thus the cross-section is linear in $s^\mu$
\begin{equation}
\label{eq:invM2}
|\amp|^2 = a + b_\mu s^\mu\,,
\end{equation}
where $b_\mu$ contains all the information about the polarization of the
top in the process. The parton level cross section can be written as
\begin{equation}
d \sigma = \Phi_{\rm in} |\amp|^2 d\Gamma_{th},
\end{equation}
where $\Phi_{\rm in}$ is the initial state flux normalization and
$d\Gamma_{th}$ is the $th$ phase space volume. On the other
hand, in the top rest frame it is convenient to introduce the spin
density matrix as
\begin{equation}
\label{eq:Rho}
\rho = A  + B_i \sigma_i,
\end{equation}
such that the unpolarized cross section is proportional to
$\overline{|\amp|^2} = \mrm{Tr}[\rho] = 2A$.  Here $\bm{\sigma}$ are
the Pauli matrices. In the density matrix formalism, the expectation
value of a generic operator is obtained as
\begin{equation}
\Braket{\mc{O}} = \mathrm{Tr} \left[\rho \mc{O} \right].
\end{equation}
In particular, the polarized cross section along $\hvec{s}$ is obtained as the
expectation value of the projector:
\begin{equation}
\label{eq:polAmp}
|\amp|^2 = \mathrm{Tr} \left[\rho\,\frac{1}{2} (1+\hvec{s} \cdot \bm{\sigma})\right] = A + B_i \hat{s}_i.
\end{equation}
One can determine the rest-frame coefficients $A, B_i$ from $a$,
$b_\mu$ by comparing the expressions for polarized $|\amp|^2$, expressed via
Eq.~\eqref{eq:invM2} and Eq.~\eqref{eq:polAmp}. The result of this matching are explicit expressions:
\begin{equation}
\label{eq:ABfromab}
A = a,\qquad B_i = -b^i + \frac{1}{m_t}\left(b^0-\frac{\bm{b}\cdot \bm{k}}{E_t + m_t}\right) \bm{k}.
\end{equation}
The rest-frame polarization of the top quark along a vector $\hvec{s}$ is given by the expectation value of $\mc{O}_{\hvec{s}} = \hvec{s}\cdot \frac{\bm{\sigma}}{2}$,
\begin{equation}
\Braket{\mc{O}_{\hvec{s}}} = \bm{B}\cdot \hvec{s}.
\end{equation}
This observable can be determined for example by measuring the angular distribution of the charged lepton in the
semi-leptonic top decay ($t \to b (W \to \ell \nu)$)\footnote{The charged lepton in top decay is considered to be an almost perfect top spin
	analyzer --  i.e. the angular decay distribution vanishes when the lepton momentum is opposite to the spin of $t$~\cite{Atwood:2000tu}.}  thus allowing for experimental
extraction of the $B^i$ coefficients~\cite{Atwood:2000tu}:
\begin{equation}
\frac{1}{\Gamma_t} \frac{d\Gamma_t}{d\cos\theta_\ell} = \frac{1}{2} \left(1+\Braket{\mc{O}_{\hvec{s}}} \cos \theta_\ell\right).
\end{equation}
Here $\theta_\ell$ is an angle between the lepton and the polarization
axis $\hvec{s}$ in the top rest frame. The above construction shows
that the vector $\hvec{s}$ is an arbitrary unit vector defined in the
laboratory frame. A particular choice $\hvec{s} = \bm{k}/|\bm{k}|$
implies that $\Braket{\mc{O}_{\hvec{s}}}$ measures the top quark
helicity. Another natural choice for $\hvec{s}$ is the $W$ momentum
$\hvec{p}$, also known as the beam basis, which has to be redefined in
$pp$ collisions where the $W$ momentum has a discrete ambiguity.
Experimentally one has to reconstruct the top quark rest frame in
order to be able to trace the angular distribution of the lepton with
respect to the chosen $\hvec{s}$ and gain access to the coefficients
$B_i$. In the following we will optimize the choice of $\hvec{s}$ such
that the sensitivity to the $CP$-violating parameter $\tilde \kappa$
is maximized.

In the $W b$ center-of-mass frame we can define the $W$ and $t$ momenta as
\begin{equation}
\label{eq:coordinate_system}
\begin{split}
\hvec{p} &= (0,0,1), \\
\hvec{k} &= (\sin\theta,0,\cos\theta),
\end{split}
\end{equation}
where $\theta$ is the angle between the direction of the top quark and
the $W$ boson. We have set the azimuthal angle $\phi=0$ without loss
of generality. The polarization vector components $B_i$ in this
case depend on $x=\cos\theta$, and we have found that in the
coordinate system in Eq.~\eqref{eq:coordinate_system} the analytical
expression for $B_2(x)$ is linear in $\tilde{\kappa}$, i.e.,
$B_2(x) = \beta(x) \tilde\kappa$, whereas $B_{1,3}$ do not contain linear $\tilde \kappa$ terms.
Effectively this means that we should
choose the vector $\hvec{s}$ to be be orthogonal to the plane spanned by
the $W$ and $t$ momenta in order to probe $\tilde \kappa$ with linear
sensitivity. Similar results have been found in Ref.~\cite{1807.00281}. We fix
$\hvec{s} = \hvec{p} \times \hvec{k}/|\hvec{p} \times \hvec{k}|$. In
this case the interesting experimental quantity is the following
two-fold differential cross-section

\begin{equation}
\label{eq:d2sigma}
\begin{split}
\frac{d^2\sigma}{dx\,d\cos \theta_\ell}(Wb \to h b \ell \nu) &=  \Sigma(x,\hvec{s}) \frac{\Br(t\to b \ell \nu)}{2}(1+\cos\theta_\ell) \\
&\phantom{=}+\Sigma(x,-\hvec{s})\frac{\Br(t\to b \ell \nu)}{2}(1-\cos\theta_\ell)\,,
\end{split}
\end{equation}
where we have approximated the intermediate top quark as a narrow resonance and $\Sigma(x,\hvec{s}) = d\sigma/dx(Wb \to t^{(\hvec{s})}h)$ is the differential production cross section for the top quarks polarized in the $\hvec{s}$ direction. Using Eq.~(\ref{eq:polAmp}) and inserting $\hvec{s}$ we have $\Sigma(x,\pm\hvec{s}) = \Phi_{\rm in}( A(x)\pm \tilde{\kappa} \beta(x))$, where $\Phi_{\rm in}$ is the initial flux normalization. Thus we can write Eq.~\eqref{eq:d2sigma} as
\begin{equation}
\label{eq:d2sigmaExplicit}
\frac{d^2\sigma}{dx\,d\cos \theta_\ell}(Wb \to h b \ell \nu) = \Phi_{\rm in} \Br(t\to b \ell \nu) (A(x)+\tilde{\kappa} \beta(x)\cos\theta_\ell)\,.
\end{equation}

\begin{figure}[t]
	\centering\includegraphics[width=0.6\textwidth]{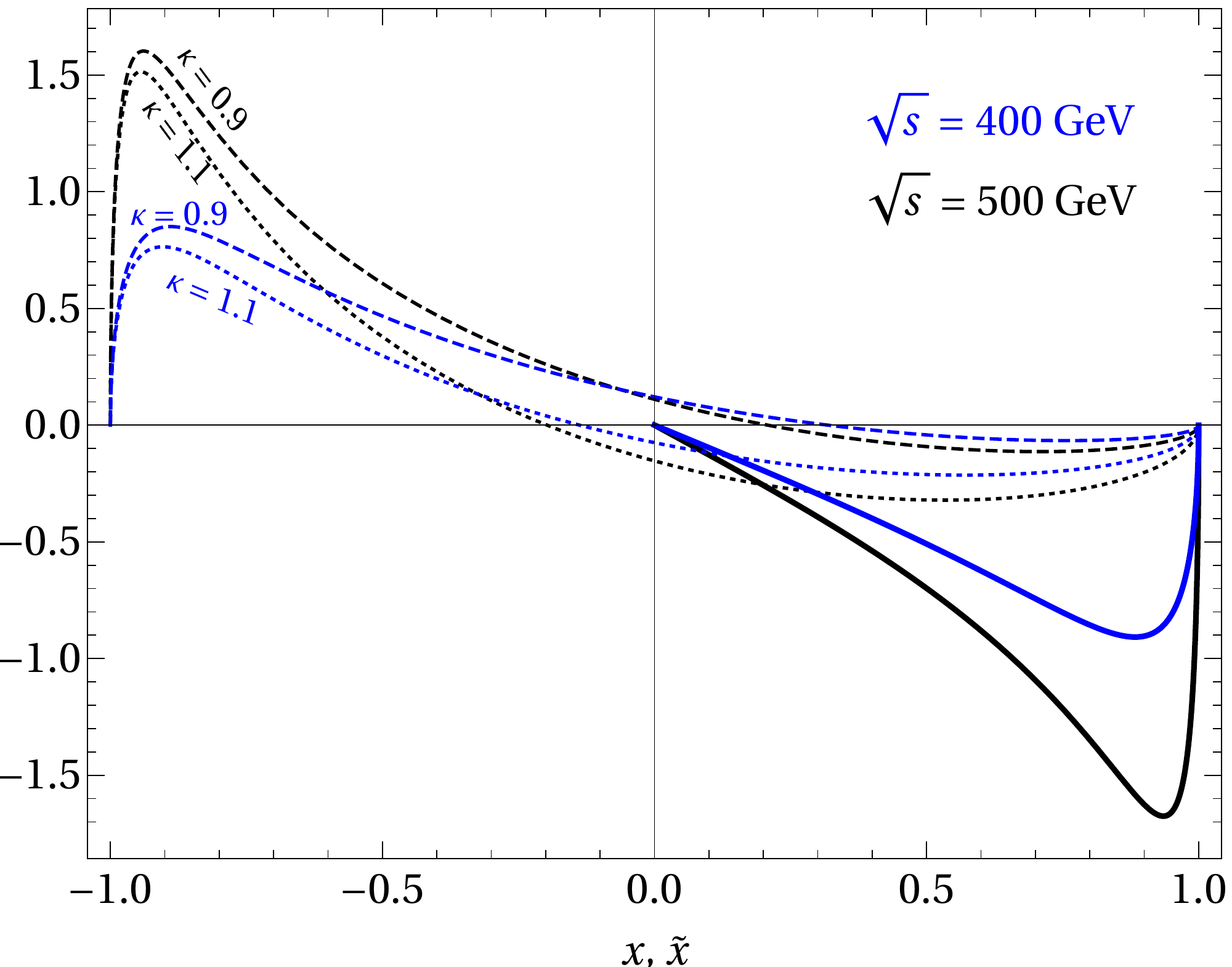}
	\caption{Comparison of the $\beta(x)$ (dashed and dotted) and $\tilde \beta(\tilde x)$ (full line) polarization functions at representative CMS energies $\sqrt{s}$ and two values of $\kappa$. We find that $\tilde \beta(\tilde x)$ is independent of $\kappa$.}
	\label{fig:B2signswitch}
\end{figure}

Treating $\tilde \kappa$ as a small perturbation we can integrate the distribution in Eq.~\eqref{eq:d2sigmaExplicit} with a phase-space dependent function $f$ that would maximize statistical sensitivity of the integral to $\tilde \kappa$. It has been shown in Refs.~\cite{Atwood:1991ka,hep-ph/9605326} that such an optimal function should be the ratio of the $\tilde \kappa$-perturbation to the unperturbed distribution, in our case $f(x,\cos\theta_\ell) = \frac{\beta(x)}{A(x)} \cos\theta_\ell$. The optimal observable is thus
\begin{equation}
\label{eq:OptParton}
\mc{O}^{Wb \to th}_\mathrm{opt.} \equiv \frac{1}{\sigma}\int dx\,d\cos\theta_\ell   \frac{d^2\sigma}{dx\,d\cos \theta_\ell} \frac{\beta(x)}{A(x)} \cos\theta_\ell = \frac{1}{N}\sum_{i=1}^N\frac{\beta(x_i)}{A(x_i)} \cos\theta_{\ell,i},
\end{equation}
where $\theta_\ell$ is the angle between $\hvec{s}$ and the lepton momentum in the top center-of mass-frame, as defined in the preceding paragraph. The index $i=1,\ldots,N$ labels individual events. The prediction scales as $\Braket{\beta^2}$,
\begin{equation}
\label{eq:OptPartonExplicit}
\mc{O}^{Wb \to th}_\mathrm{opt.} = \frac{\tilde \kappa}{3}\, \left[\int dx\, \frac{[\beta(x)]^2}{A(x)}\right]\Big/\left[\int dx \,A(x) \right],
\end{equation}
where we have integrated over $\cos \theta_\ell$ and left the bounds for $x = \cos \theta$ unspecified. The function $\beta(x)$ is plotted in Fig.~\ref{fig:B2signswitch}.

To carry over the presented formalism to the realistic case of $pp$
collisions, we have to adapt the beam axis by referring only to
experimentally accessible momenta. Using the reconstructed top
momentum $\bm{k}$ as a reference, we define the positive $z$-direction
as the parallel top quark momentum projection $\hvec{k}_{\parallel}$. The top quark is then always in the positive
hemisphere, $\tilde x = \cos\tilde\theta \geq 0$, where $\tilde\theta$
is the angle between $\bm{k}$ and $\hvec{k}_\parallel$. The polarization direction with linear $\tilde \kappa$ sensitivity now becomes $\hvec{s} = \hvec{k}_\parallel \times \hvec{k}_\perp$ upon which we now measure the lepton angle $\tilde\theta_\ell$. The cross-section distributions in $\tilde x$ and $x$ are related via
\begin{equation}
\label{eq:d2sigmaTilde}
\begin{split}   
\frac{d^2\sigma}{d\tilde x\,d\cos \tilde\theta_\ell} &= \left.\frac{d^2\sigma}{dx\,d\cos \theta_\ell}\right|_{x=\tilde x,\cos \theta_\ell = \cos \tilde\theta_\ell} + \left.\frac{d^2\sigma}{dx\,d\cos \theta_\ell}\right|_{x=-\tilde x,\cos \theta_\ell = -\cos \tilde\theta_\ell}\\
&= \Phi_{\rm in} \frac{\mathrm{Br}(t\to b \ell \nu)}{2}\Big[\tilde A(\tilde x)  + \tilde \kappa \cos\tilde\theta_\ell \tilde \beta(\tilde x) \Big],
\end{split}
\end{equation}
where
\begin{equation}
\begin{split}       
\tilde A(\tilde x) &\equiv A(\tilde x) + A(-\tilde x)\,,\\
\tilde \beta(\tilde x) &\equiv  \beta(\tilde x) - \beta(-\tilde x).
\end{split}
\end{equation}  
The $\cos\theta_\ell$ is flipped in the second term since for 
$\tilde x = -x$ the polarization vector $\hvec{s} = \hvec{k}_\parallel \times \hvec{k}_\perp$ flips the direction compared to the previous definition, $\hvec{s} \sim \bm{p} \times \bm{k}$. The optimal observable in this case
is finally
\begin{equation}
\label{eq:OptPartonTilde}
\begin{split}      
\tilde{\mc{O}}^{Wb \to th}_\mathrm{opt.} &\equiv \frac{1}{\sigma}\int d\tilde x\,d\cos\tilde\theta_\ell   \frac{d^2\sigma}{d \tilde x\,d\cos \tilde\theta_\ell} \cos\tilde\theta_\ell\, \frac{\tilde{\beta}(\tilde x)}{\tilde{A}(\tilde{x})} = \frac{1}{N}\sum_{i=1}^N\frac{\tilde \beta(\tilde x_i)}{\tilde A(\tilde x_i)} \cos\tilde\theta_{\ell,i},\\
&=\frac{\tilde \kappa}{3}\,\left[\int d\tilde x\, \frac{[\tilde \beta(\tilde x)]^2}{\tilde{A}(\tilde{x})}\right]\Big/\left[\int d \tilde x \,\tilde A(\tilde x) \right].
\end{split}
\end{equation}
In the limit where $\beta(x) = - \beta(-x)$ the observables are equal,
$\tilde O_\mathrm{opt.}^{Wb \to th} = O_\mathrm{opt.}^{Wb \to th}$. However in general the $\tilde O_\mathrm{opt.}^{Wb \to th}$ is expected to result in a weaker statistical significance due to our inability to determine the direction of the top quark with respect to the initial $W$. Fig.~(\ref{fig:B2signswitch}) shows that  $\beta(x)$ is large at negative $x$ and we have $\tilde\beta(\tilde x) \approx -\beta(-\tilde x)$, for representative values of the center-of-mass energy $\sqrt{s}$.

\subsection{Hadronic process $pp \to t h j$}
\begin{figure}[!b]
	\centering
	\begin{tabular}{c}
		\includegraphics[width=0.45\linewidth]{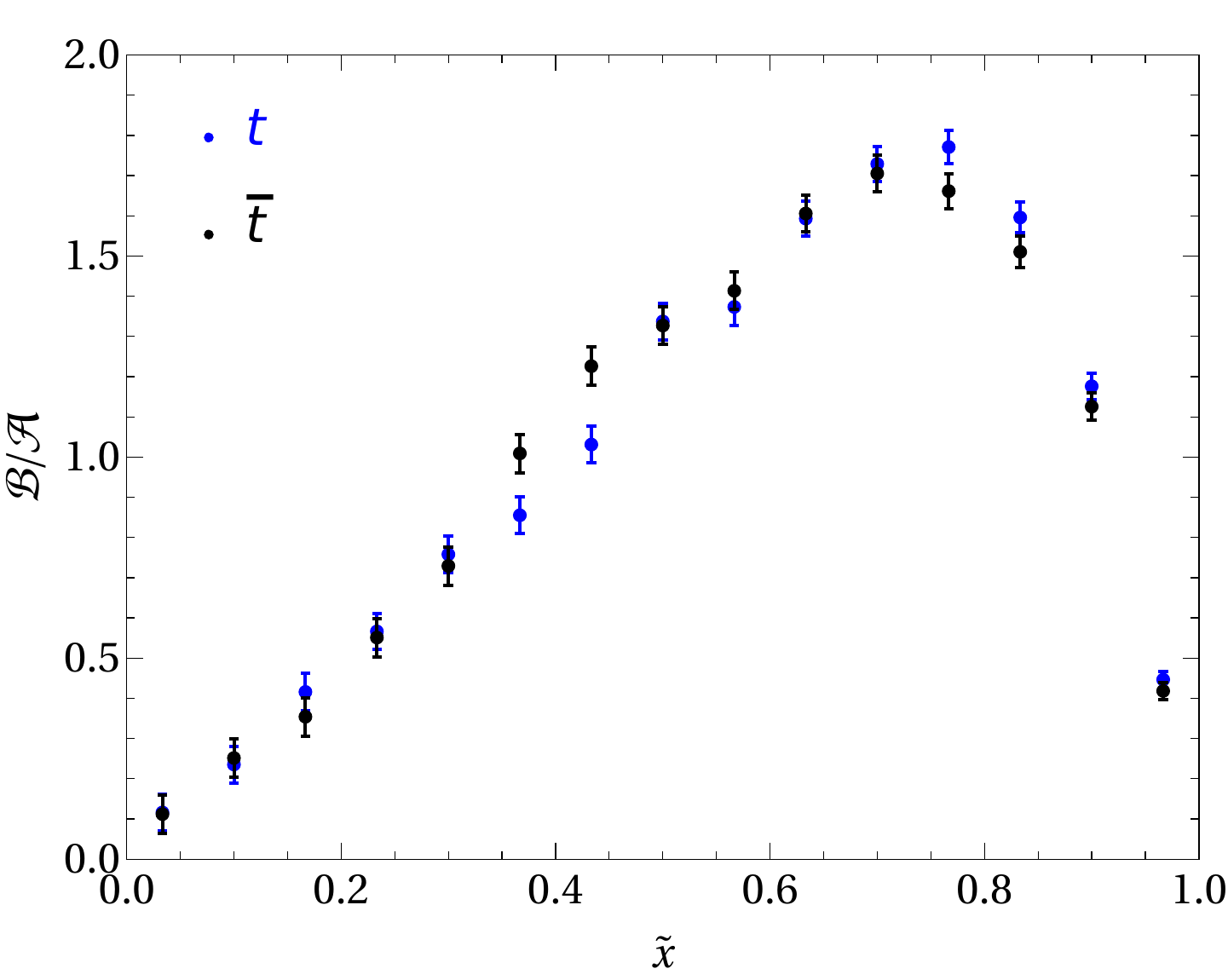}\ \ \includegraphics[width=0.45\linewidth]{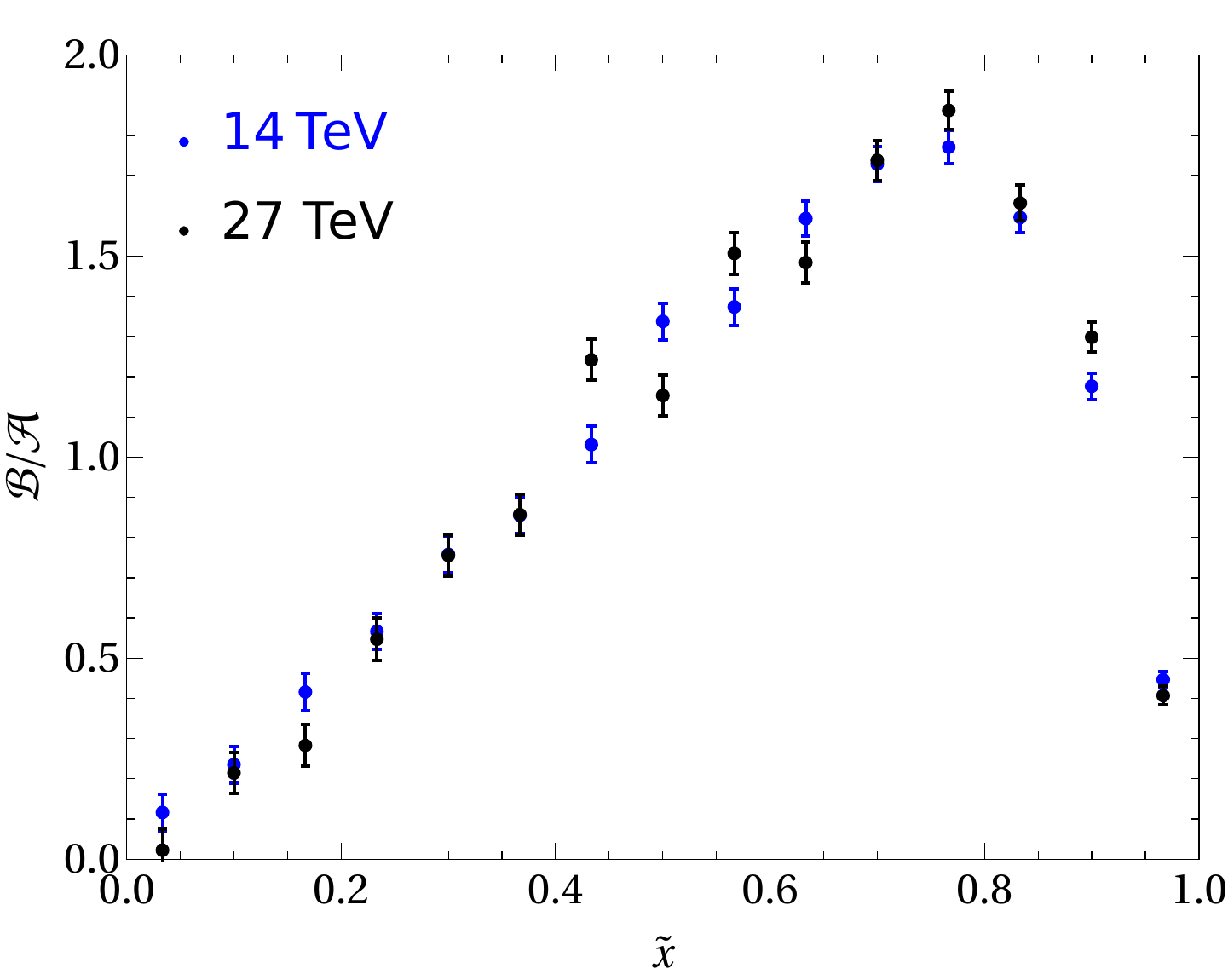}
	\end{tabular}
	\caption{Comparison of the optimal weight $\mc{B}/\mc{A}$ between the $pp \to t h j$ and $pp \to \bar t h j$ processes extracted from MC simulations (left). The right panel shows the comparison between $14$ and $27\e{TeV}$ proton collision energies for $pp \to thj$ . All plots are obtained using $\tilde{\kappa}=1$ and with $10^6$ MC events.}
	\label{fig:extracted_beta_2}
\end{figure}
\begin{figure}[!h]
  \centering
  \begin{tabular}{lr}
    \includegraphics[width=0.45\linewidth]{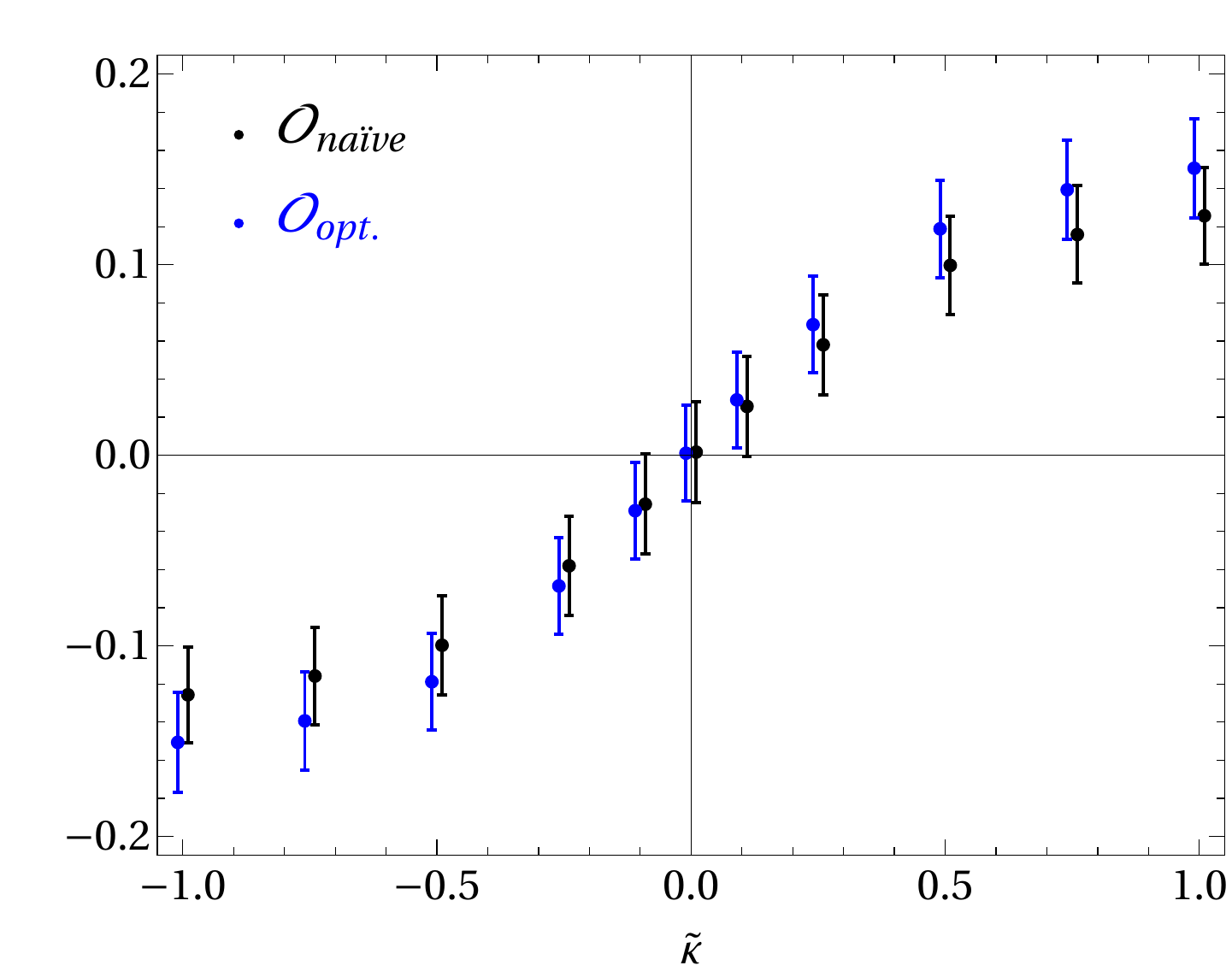}\ \  \ \includegraphics[width=0.43\linewidth]{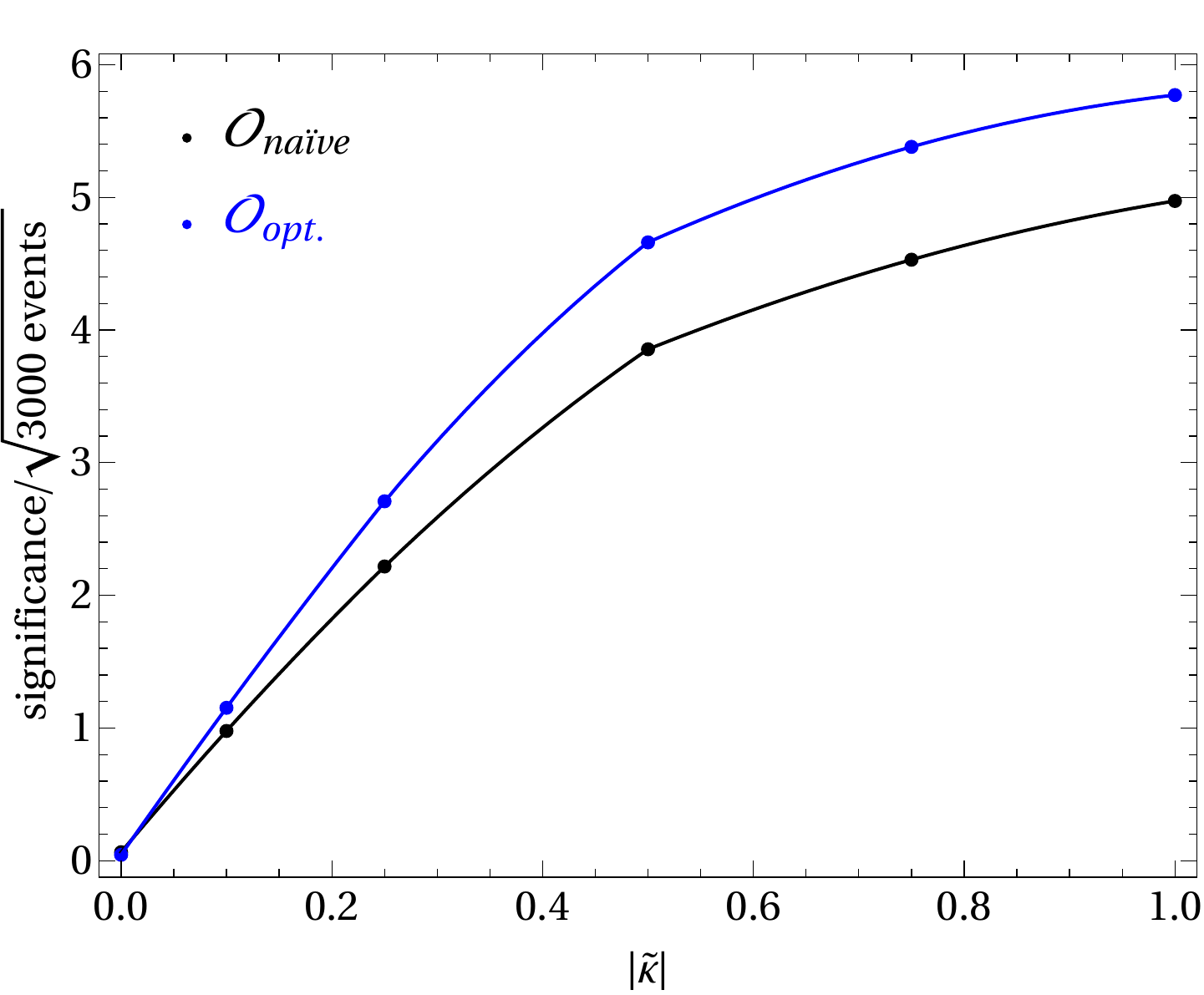}
  \end{tabular}
	
	\caption{Left: comparison of the optimized spin observable (blue dots) with the na\"ive observable (black dots) extracted from $3000$ $p p \to t h j, t \to b \ell \nu$ MC events at each choice of $\tilde \kappa$. Right: comparison of the significance (defined as the mean value divided by the standard deviation) per $\sqrt{N}$ of the two observables, where $N$ is the number of events.}
	\label{fig:pp-thj-kptld}
\end{figure}

Here we demonstrate the procedure of measuring the optimal observable in the case of $pp$ collisions, but still neglecting reconstruction efficiencies and backgrounds. The parton level observable
defined in Eq.~\eqref{eq:OptPartonTilde} can be adapted to this case with an
additional integration over the parton distribution functions
(PDFs). Since the hadronic cross section is a convolution of partonic
cross sections it can be split into a $\tilde \kappa$-independent piece and the
small perturbation proportional to $\tilde \kappa$, similar to the partonic cross section in Eq.~\eqref{eq:d2sigmaTilde}. Assuming that the
Higgs decays into visible states, the missing $p_T$ is only due to the
neutrino originating from the top decay. Thus we can reconstruct the top quark momentum and
kinematic quantities of
Eq.~\eqref{eq:d2sigmaTilde}. Thus, for hadronic collisions one can
express the cross section as
\begin{equation}
\label{eq:HadronicThj}  
\frac{d^2\sigma^{pp\to thj }}{d\tilde x\,d\cos \tilde\theta_\ell} = \mc{A}(\tilde x) + \tilde \kappa \cos\tilde \theta_\ell \mc{B} (\tilde x) ,
\end{equation}
and weigh the events with the optimal
$f_\mrm{opt.}  \propto \cos\tilde\theta_\ell \mc{B}/ \mc{A}$. We use the MC event generator
$\texttt{MadGraph5}$~\cite{Alwall:2014hca,1212.3460} together with the Higgs
Characterisation UFO model~\cite{Degrande:2011ua, Artoisenet:2013puc} (for an analysis of NLO QCD and NNLL EW effects see  Refs.~\cite{1407.5089, 1504.00611} and~\cite{1907.04343}, respectively) to incorporate the
$\kappa$ and $\tilde \kappa$ couplings in the simulation of the
$p p \to t (\to b \ell \nu) h j$ signal. The procedure of extracting the weight function $\mc{B}/\mc{A}$ from MC
simulations and using it to produce the optimal observable goes as
follows:
\begin{enumerate}
	\item Choose the bins for $\tilde{x}$ between $\tilde x_\mrm{min} \ge 0$ and $\tilde x_\mrm{max} \le 1$.
	\item Fix $\tilde{\kappa}$ and extract from the MC simulation the mean $\langle\cos\tilde{\theta}_\ell\rangle$ in each of the $\tilde x$ bins. The obtained value corresponds to weight $\frac{1}{3}\mc{B}/\mc{A}$ in this bin, see Eq.~\eqref{eq:HadronicThj}.
	\item Use this information to weigh experimental events bin-by-bin with $f_\mrm{opt.} \propto \cos\tilde\theta_\ell \mc{B}/ \mc{A}$. The normalization of $f_\mrm{opt.}$ is fixed by the requirement $\int d\tilde x \mc{B}/ \mc{A} = 1$.         
\end{enumerate}
This optimization procedure is independent of the $\tilde \kappa$ value. The resulting optimal weight $\mc{B}/\mc{A}$ is shown in Fig.~\ref{fig:extracted_beta_2}, where we compare it for different final states ($t h j$ or $\bar t h j$) and collision energies ($14$ or $27\e{TeV}$). We have also extracted the weight function from simulations at NLO in QCD to estimate the systematic uncertainty associated with higher order QCD effects and found that the difference is within 10$\%$ of the LO extraction. Finally, we compare our optimized approach to the na\"ive $\tilde \kappa$ extraction through the measurement of $\Braket{\cos \theta_\ell}$, which in turn corresponds to the case where the weight is independent of $\tilde x$, i.e. $\mc{B}/\mc{A} = 1$.
Fig.~\ref{fig:pp-thj-kptld} shows the improvement of the significance when the optimal weight function is applied on simulated signal events without showering or reconstruction effects at 14 TeV.

\subsection{Limits in the ($\kappa$, $\tilde{\kappa}$) plane from $pp\to thj$ at event reconstruction level}

In order to make closer contact with experiments, we now include the effects of parton showering, detector response and background processes. We use $\texttt{MadGraph5}$ to generate events at leading order (LO) in QCD for the signal process $p p \to t (\to b \ell \nu) h (\to b \bar{b}) j$ plus the conjugate process with $\bar t$ at  14~TeV High-Luminosity LHC (HL-LHC) and 27~TeV High-Energy LHC (HE-LHC) center-of-mass energies.\footnote{Note that our procedure of obtaining an optimal observable does not depend on the $h$ decay products, therefore this analysis should be taken as a proof of concept with potential for future improvements using e.g.~multiple $h$ decay channels.} Event generation is performed for multiple values of ($\kappa$, $\tilde{\kappa})$. The parton level events are subsequently showered and hadronizied with $\texttt{Pythia8}$~\cite{Sjostrand:2007gs}, and jets are clustered with the anti-$k_T$ algorithm using $\texttt{FastJet}$~\cite{Cacciari:2011ma}. For detector simulation and final state object reconstruction (e.g. lepton isolation and $b$-tagging) we use $\texttt{Delphes v3.3.3}$~\cite{deFavereau:2013fsa} with the default ATLAS parameters in $\texttt{delphes\_card\_ATLAS.tcl}$. The dominant background process in this analysis is $t\bar t$ production with additional associated jets. We include this background by generating $pp\to t\bar t$ samples, with one of the tops decayed into the semi-leptonic channel and the other one decayed into the hadronic channel, produced in association with $0$, $1$ and $2$ hard jets. In order to correctly model the hard jets' distributions, we merge the matrix element computations with the MC shower using the $\texttt{MLM}$~\cite{Mangano:2006rw} prescription. 
For the event selection we demand the following basic requirements:
\begin{itemize}
\item Exactly $3$ $b$-tagged jets with $|\eta(b)|<5$ and $p_T(b)>20$~GeV,
\item One additional (non-tagged) light jet exclusively in the forward direction with $2<|\eta(j)|<5$ and $p_T(j)>20$~GeV,
\item One isolated light lepton $\ell^\pm=e^\pm,\mu^\pm$ with $|\eta(\ell)|<2.5$ and $p_T(\ell)>10$~GeV.  
\end{itemize}

\noindent In addition, we further select events with one reconstructed Higgs and one reconstructed top quark as follows: first, we calculate the three possible invariant masses from the three reconstructed $b$-jets ($m_{bb}$) and only keep the event if at least one $bb$ pair satisfies $|m_H-m_{bb}|<15$~GeV. For such events, we select as the Higgs decay candidate $h\to b\bar b$ for the pair of $b$-jets with the invariant mass closest to the Higgs mass. The remaining non-Higgs $b$-jet is then assumed to come from the top-quark decay. Next, we reconstruct the top-quark by requiring that the combined invariant mass $m_{bl\nu}$ of the remaining $b$-jet, the lepton, and the neutrino (also reconstructed by assuming it to be the unique source of missing energy in the event) to fall inside the mass window of the top-quark defined by $m_t\pm 35$~GeV. In order to further reject the $t
\bar t$ backgrounds, events with a reconstructed Higgs and top are selected if the combined invariant mass of the $b$-jets originating from the Higgs and the light jet satisfies the cut $m_{bbj}>280$~GeV \cite{Farina:2012xp}. The final selection efficiency for the $thj$ signal in the SM is $0.32\%$ ($0.23\%$), while for the background it is $0.008\%$ ($0.006\%$) at 14 TeV (27 TeV).\\

\begin{figure}[t!]
	\centering 	
	\includegraphics[width=0.5\linewidth]{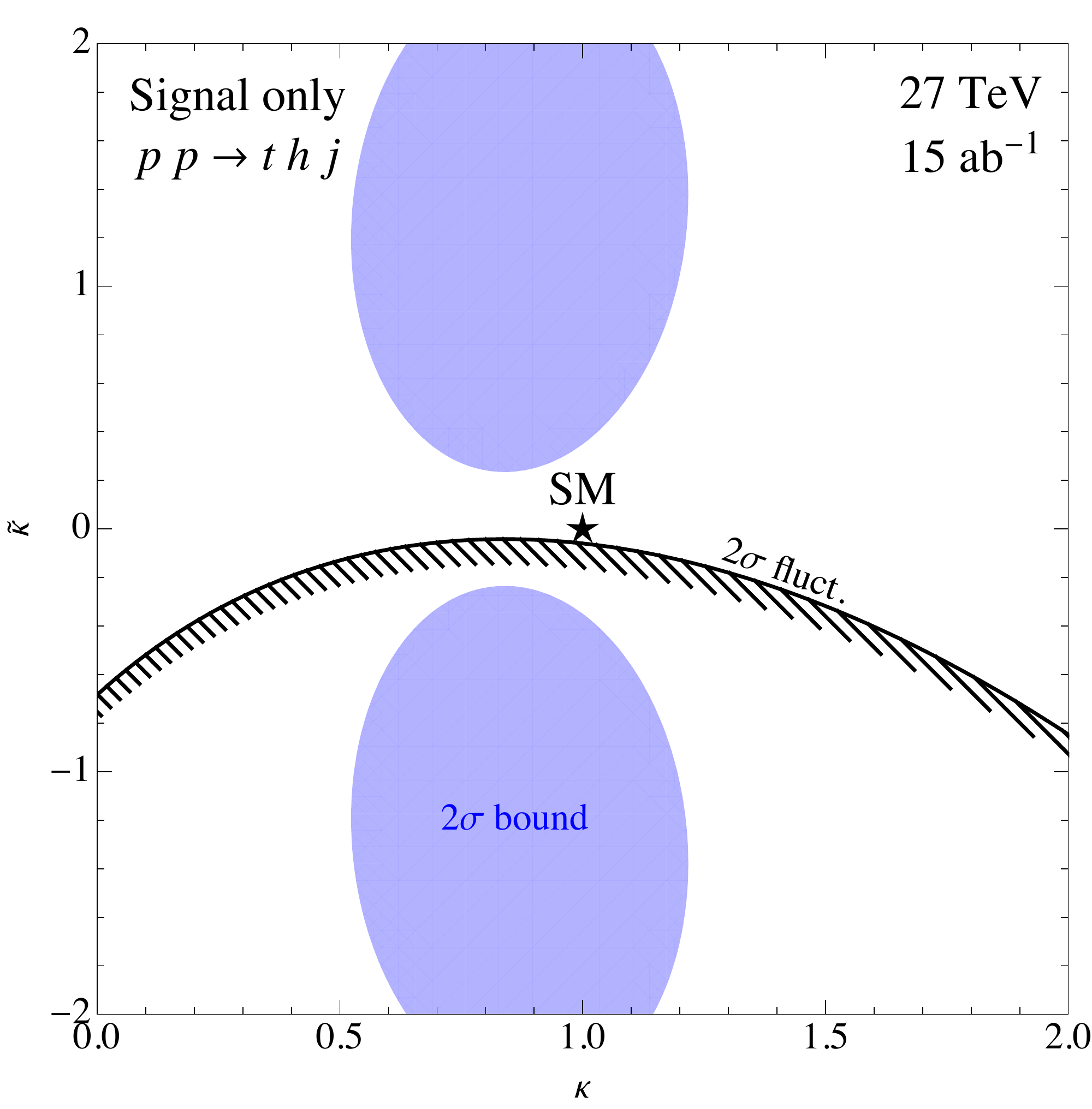}
	\caption{Bounds in the $(\kappa, \tilde{\kappa})$ plane using the optimized observable $\mc{O}_\mathrm{opt}$ for the single-top associated production with a Higgs boson. The blue shaded region corresponds to the $2\sigma$ ($\chi^2 > 6.18$) exclusion zone assuming the measurement of the SM at the HE-LHC (15 ab$^{-1}$). The black line and stripes shows the $2\sigma$ excluded region for a $2\sigma$ positive fluctuation at the HE-LHC (see text for details).}
	\label{fig:th-HE-LHC}
\end{figure}

As we fully reconstruct the $th$ system and have access to the lepton momentum from the top decay we have all the necessary information for measuring the optimized spin observable. We use the optimal weight function $\mc{B}/\mc{A}$ (Fig.~\ref{fig:extracted_beta_2}) extracted from the MC simulations to construct a $\chi^2$ with an appropriately weighted signal process. Our results for $pp\to thj$ generated in the SM are given by the $2\sigma$ exclusion limits (shaded blue) shown in Fig.~\ref{fig:th-HE-LHC} for the HE-LHC at a luminosity of $15\,\text{ab}^{-1}$. As can be seen in Eq.~\eqref{eq:OptPartonTilde} the observable $\mc{O}_\mathrm{opt.}$ is normalized to the cross section, which contains terms $\kappa^2$, $\tilde \kappa^2$, as well as a linear term in $\kappa$ and a constant term due to second diagram in Fig.~\ref{fig:diagramsWbth}, whereas the numerator $\propto \tilde\kappa (\kappa+c)$. The behaviour of $\mc{O}_\mathrm{opt}$ close to the SM point is thus linear in $\tilde \kappa$, whereas the cross section has a minimum in $\kappa$ close to $\kappa=1$. In the large coupling regime $\mc{O}_\mathrm{opt.}$ is converges to a small value which depends on the direction in which we make the limit $\kappa^2 +  \tilde \kappa^2 \to \infty$. The $2\sigma$ exclusion has an elliptic shape, but according to the presented analysis, milder exclusion regions would have hyperbolic shapes.
We also present the ellpitic limit (given by the black elliptic contour) assuming a $2\sigma$ positive excess above the SM expectation corresponding to a measurement of the optimized spin observable of $O_{\text{opt.}}=0.06 \pm 0.03$ whose size and error are statistics-driven. Because of the nature of our observable, the signed fluctuation gives rise to asymmetric limits in the $\tilde\kappa$ direction. In the $\kappa$ direction the bounds are also not symmetric as $pp\to thj$ production is sensitive to $\text{sgn}(\kappa)$. 
Finally, in order to include background effects, the same statistical analysis would have to be repeated including the $t\bar t$ background in the $\chi^2$ fit. However, even with a large background rejection as implemented above, the irreducible background is simply too large and the signal is completely diluted leading to a signal significance of only $S/\sqrt{B} \sim 0.8~(3.2)$ at 14 TeV (27 TeV) at a luminosity of $3\,\text{ab}^{-1}$ ($15\,\text{ab}^{-1}$). This effectively precludes any meaningful extraction of bounds on $\tilde \kappa$ from a fit to $O_{\rm opt.}$. We leave the possibility of further optimizing the cuts in order to reduce the backgrounds or including other Higgs decay channels for future works. In the following we instead focus on the related but more abundant process of associated top quark pair and Higgs boson production.

%
\section{$CP$-odd observables in $pp \to t \bar{t} h$}\label{sec:tth}
%
In this section we consider new $CP$-odd observables in the process $p p \to t \bar{t} h$, with both top quarks decaying semi-leptonically. Compared to $p p \to t  h j$, this process has a much better S/B ratio and has in fact been recently measured by the LHC collaborations~\cite{1712.08895, 1806.00425}.\footnote{For the state of the art predictions of the differential distributions see e.g. Ref.~\cite{1907.04343}.} 
The top quarks in this process are known to be unpolarized, independent of the $\tilde \kappa$ value~\cite{Ellis:2013yxa}. Information on the underlying $\kappa$ and $\tilde\kappa$ parameters is nonetheless contained in the correlations among the top spins. Direct experimental extraction of top polarizations in $p p \to t \bar{t} h$ suffers from combinatorial difficulties with reconstructing both $t$ and $\bar t$ rest frames. Therefore in the following we focus directly on lab frame kinematic distributions in variables which are $CP$- and $P$-odd and are constructed from accessible final-state momenta~\cite{Boudjema:2015nda}.

\subsection{Laboratory frame $CP$-odd observables}
\label{sec:labFrameTTH}
 We denote the $3$-momenta of the leptons and $b$-jets originating from $t$ and $\bar{t}$ with $\lp$, $\lm$, $\b$ and $\bbar$, respectively, and the Higgs $3$-momentum with $\h$. The $C$ and $P$ transformation properties of six independent combinations of these momenta are given in Tab.~\ref{tab:CP_transformations_3_momenta}. We focus only on combinations that are nontrivial under $C$, $P$ (i.e., we omit scalars products) and are accessible in a realistic experimental environment. For example we consider $\b + \bbar$, but not $\b - \bbar$ as differentiating between $b$ and $\bar{b}$ is difficult experimentally.\footnote{For recent attempts in extracting the charge of the $b$-jet see Refs.~\cite{Krohn:2012fg, Fraser:2018ieu,ATLAS-2015-040,ATLAS:2018lhe}.}
\begin{table}[t!]
\begin{center}
	\begin{tabular}{c||c|c|c|c|c|c}		
          & $\h$ & $\lm+\lp$ & $\lm-\lp$ & $\b + \bbar$ & $\lm\times\lp$ & $\b \times \bbar\,(\b-\bbar)$\\
          \hline\hline
          $C$ & $+$ & $+$ & $-$ & $+$ & $-$ & $+$\\

          $P$ & $-$ & $-$ & $-$ & $-$ & $+$ & $-$ \\
          \hline\hline
          $CP$ & $-$ & $-$ & $+$ & $-$ & $-$& $-$ \\ 		
	\end{tabular} 
\end{center}
\caption{Momenta with well-defined $C$ and $P$ eigenvalues. The $b,\bar b, \ell^+$ and $\ell^-$ are the top decay products. The last column is a rank-2 tensor -- a direct product of an axial and a polar vector.}
\label{tab:CP_transformations_3_momenta}
\end{table}
The six combinations of momenta in Tab.~\ref{tab:CP_transformations_3_momenta} are taken as a basis for
constructing $P$- and $CP$-odd variables $\omega$. This is achieved
by contracting (anti)symmetrically the momentum tensors such that the resulting
$\omega$ is $C$ even and $P$ odd. i.e. a pseudoscalar. The resulting spectrum is then linear in the pseudoscalar $\omega$ with the coefficient in front linear in $\tilde \kappa$, analogous to expression~\eqref{eq:HadronicThj}. At leading order in $\tilde \kappa$ we find:
\begin{equation}
\label{eq:tthSpectrum}
\frac{d^2\sigma}{d x d \omega} = \mc{C}(x) + \kappa \tilde{\kappa} \mc{D}(x) \omega .
\end{equation}
In Eq.~\eqref{eq:tthSpectrum} we have parameterized the phase space
with the pseudoscalar variable $\omega$, while all other variables are collectively 
denoted by $x$. Now we can again extract
$\tilde \kappa$ with the statistically optimal weight function, which in this
case is given by $f_\mrm{opt.} \propto  \mc{D}(x) /\mc{C}(x)$, while the associated
observable is 
\begin{equation}
  \label{eq:Oomega}
\mc{O}_\omega = \frac{1}{\sigma} \int dx\,d\omega \frac{d^2\sigma}{dx
  d\omega} f(x) \omega  = \frac{1}{N}\sum_{i=1}^N f(x^{(i)}) \omega^{(i)}.
\end{equation}
Here $N$ is the number of experimental events. In contrast to the extraction of $f$ for the $th$
process, here the extraction of $\mc{D}/\mc{C}$ turns out to
be more complicated due to the high-dimensionality of the phase space ($4$
variables for each $p p \to t \bar t h$, $t \to b \ell^+ \nu$,
$\bar t \to \bar b \ell^- \bar \nu$). One could use a Monte Carlo event generator to obtain events following the distribution in Eq.~\eqref{eq:tthSpectrum} in order to extract $\mc{D}/\mc{C}$. However, since binning in all
dimensions is not feasible it is better to formulate the task as a maximization problem to obtain the unknown weight function $f(x;\alpha)$. Given the $N$ events $(x^{(i)}, \omega^{(i)})$, $i=1,\ldots, N$, generated with a
non-zero $\tilde \kappa$, the corresponding observable and its associated standard
deviation are obtained as
\begin{equation}
\mc{O}_\omega = \frac{1}{N} \sum_i f(x^{(i)};\alpha) \omega^{(i)}\ \ \ \ \text{and }\ \ \ \
\sigma_{\mc{O}_\omega}^2 = \frac{1}{N} \left[\frac{1}{N}\sum_i(f(x^{(i)};\alpha) \omega^{(i)})^2 - \mc{O}_\omega^2\right],
\end{equation} 
respectively. The goal is to find the set of parameters $\alpha$ of
the function $f(x; \alpha)$, defined on phase space $x$ and
parameterized by $\alpha$, that maximize the significance:
\begin{equation}
  \label{eq:tthOptimization}
  \begin{split}
    \mrm{Sig}(\alpha) &\equiv \frac{\mc{O}^2_\omega}{N \sigma^2_{\mc{O}_\omega}}= \frac{\mc{O}_\omega^2}{\frac{1}{N} \sum_i  \left[f(x^{(i)};\alpha) \omega^{(i)}\right]^2 - \mc{O}_\omega^2}.
  \end{split}
\end{equation}
The significance $\mrm{Sig}(\alpha)$ is independent of large enough
$N$.  The arguments of function $f$ could be scalar products between
final state momenta, whereas the functional form, controlled by
parameters $\alpha$, should be general enough.  The obtained $f$, that
was optimized using MC data can then be applied on a given
experimental sample. In the following we will not pursue the globally optimal
weight but will perform partial optimization along a single
dimension of phase space.\\

First we introduce the relevant $CP$- and $P$-odd variables.  The simplest pseudoscalar is a mixed product of the form
$\bm{V} \cdot \bm{A}$, where $\bm{V} (\bm{A})$ denotes vector (axial vector), an example of which is
\begin{equation}
\label{eq:bmbbar}
\mc{\omega}_{b-\bar{b}}=(\b-\bbar)\cdot (\lm \times \lp),
\end{equation}  
presented already in Refs.~\cite{ hep-ph/9312210, Boudjema:2015nda} (see also observables proposed in Ref.~\cite{1603.03632}). In our case we do not wish to use $\b -\bbar$ which leads us to an alternative mixed product that does not rely
on separating $b$ from $\bar b$ experimentally:
\begin{equation}
  \label{eq:VA}
  \omega_{h\ell b} \equiv \frac{\left[\h \times (\lm + \lp) \right]\cdot(\b +\bbar)}{|\h\times (\lm+\lp)| \,|\b+\bbar|}.
\end{equation}
Once we allow for a more complicated pseudoscalar of the form $(\bm{V} \cdot \bm{A}) \, (\bm{V}\cdot \bm{V})$
there are 13 possibilities, listed in Appendix~\ref{app:tthVars}. Out of those and the mixed product in Eq.~\eqref{eq:VA}, one variable stands out as the most sensitive one:
\begin{equation}
  \label{eq:O6}
  \omega_6 \equiv \frac{\left[(\lm \times \lp) \cdot (\b+\bbar)\right] \left[(\lm - \lp)\cdot (\b+\bbar)\right]}{|\lm \times \lp|\,|\lm-\lp| |\b+\bbar|^2}.
\end{equation}
The pseudoscalar variable $\omega_6$ is bounded\footnote{In terms of notation used to classify the variables in App.~\ref{app:tthVars} $\omega_6$ corresponds to $\omega_\ell^{b+\bar b,b+\bar b}$.} within the interval $[-1,1]$. We have found that for the differential cross section $d^2\sigma/(dx\,d\omega_6)$ (see Eq.~\eqref{eq:tthSpectrum}) the ratio $\mc{D}/\mc{C}$, where $x$ is an arbitrary kinematic variable, is approximately constant and does not oscillate in sign, which allows us to use a na\"ive weight function, $f(x) = 1$, without paying too much price for the cancellation between contributions from different regions of phase space. The observable we use is thus simply the average of $\omega_6$:
\begin{equation}
  \label{eq:Oomega6}
  \begin{split}   
  \mc{O}_6 &= \frac{1}{\sigma} \int dx\,d\omega_6 \frac{d^2\sigma}{dx
    d\omega_6} \omega_6  = \frac{1}{N}\sum_{i=1}^N \omega_6^{(i)}.
\end{split}
\end{equation}
The behavior of $\mc{O}_6$ in comparison to the analogously defined observable $\mc{O}_{b-\bar{b}}$ based on $\omega_{b-\bar{b}}$ \eqref{eq:bmbbar}, as a function of $\tilde \kappa$ is shown in Fig.~\ref{fig:tth-part}.\\

In addition to the analysis of $\mc{O}_{6}$ presented below, we have also analyzed observables related to all the other pseudoscalar variables in App.~\ref{app:tthVars}. For some of them we have used the optimization technique along a chosen dimension of phase space $x$, which in some cases drastically improved their sensitivity to $\kappa \tilde \kappa$. Nonetheless none of the other possible observables reached a sensitivity close to $\mc{O}_6$. We note that all the considered observables can be further improved in sensitivity by performing a full global phase-space optimization using Eq.~\eqref{eq:tthOptimization}. A task which we leave for future work.

\begin{figure}[t!]
	\centering
	\begin{tabular}{lr}
		\includegraphics[width=0.49\linewidth]{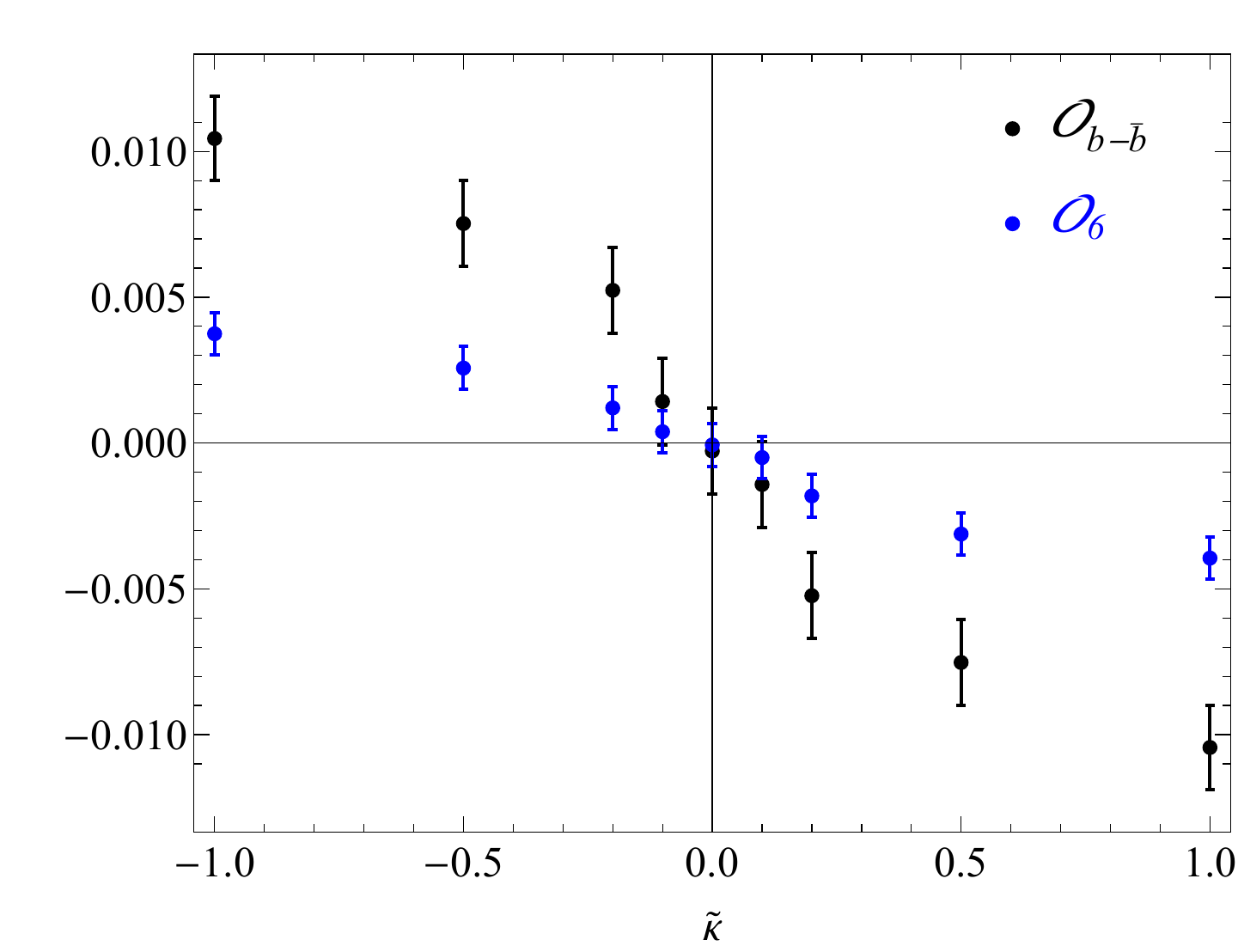}\ \  \ \includegraphics[width=0.47\linewidth]{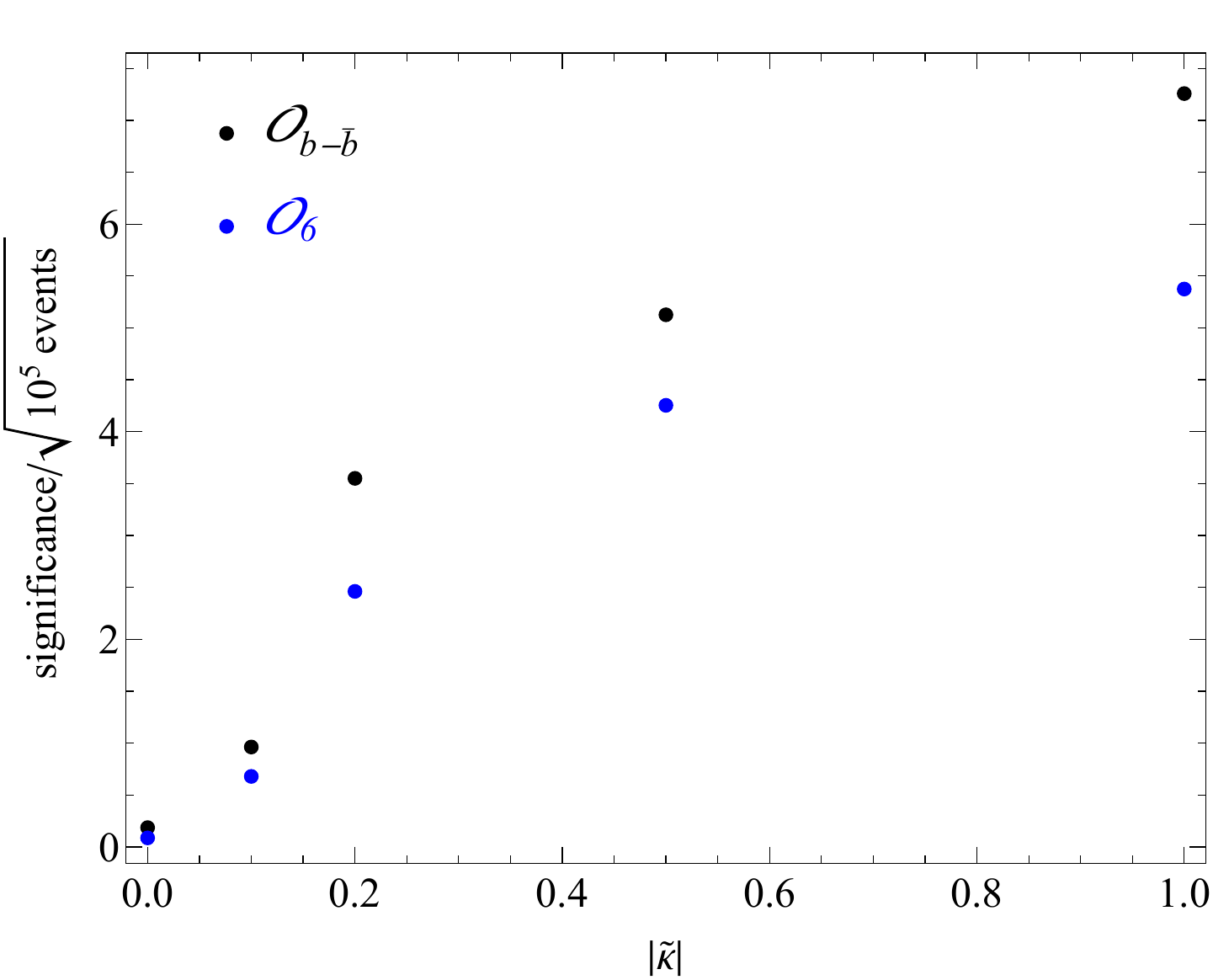}
	\end{tabular}
	
	\caption{Left: $\mc{O}_6$ \eqref{eq:Oomega6} and $\mc{O}_{b-\bar{b}}$ as functions of $\tilde \kappa$ with $\kappa = 1$ and $10^5$ events per $\tilde{\kappa}$ at 27 TeV. Right: Comparison of the significances of the same quantities. Compared to $\mc{O}_{b-\bar{b}}$ the presented $\mc{O}_6$ is slightly less significant, however the difficulties with reconstructing the $b$-jet charges are avoided in our case, rendering $\mc{O}_6$ more appealing when taking showering, hadronization and detector effects into consideration.}
	\label{fig:tth-part}
\end{figure}

\subsection{Limits in the ($\kappa$, $\tilde{\kappa}$) plane from $pp\to t\bar th$}
We now demonstrate the capability of current and future colliders to measure the $\mc{O}_6$ observable in $t\bar t h$ production. For this purpose we have generated using $\texttt{MadGraph5}$ multiple event samples of $p p \to t \bar{t} h$ for different values of ($\kappa$, $\tilde{\kappa})$, followed by the decay chain  $(t \to b \ell^+ \nu_\ell,\bar{t} \to \bar{b} \ell^- \bar{\nu}_\ell, h \to b \bar{b})$ at 14 TeV and 27 TeV. The partonic events were then fed into $\texttt{Pythia8}$ for showering and hadronization and finally into $\texttt{Delphes}$ for detector simulation with the default ATLAS card. We have followed the same steps to generate the events of the main irreducible background $p p \to t \bar{t} b \bar{b}, ( t \to b \ell^+ \nu_\ell,\bar{t} \to \bar{b} \ell^- \bar{\nu}_\ell)$. The basic event selection requirements for this analysis are:

\begin{figure}[t!]
	\centering 	
	\begin{tabular}{lr}
		\includegraphics[width=0.45\linewidth]{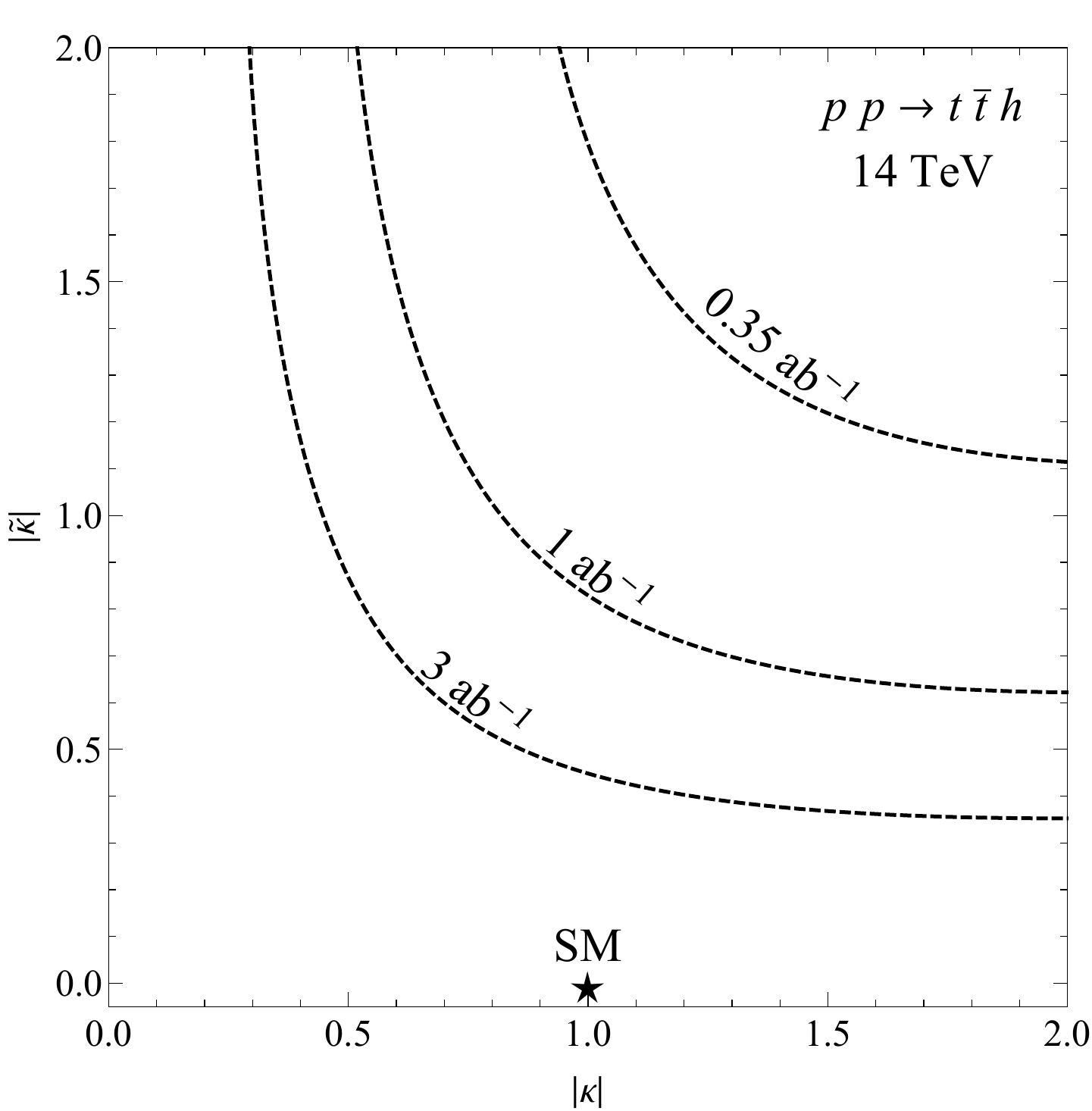} &
		\includegraphics[width=0.45\linewidth]{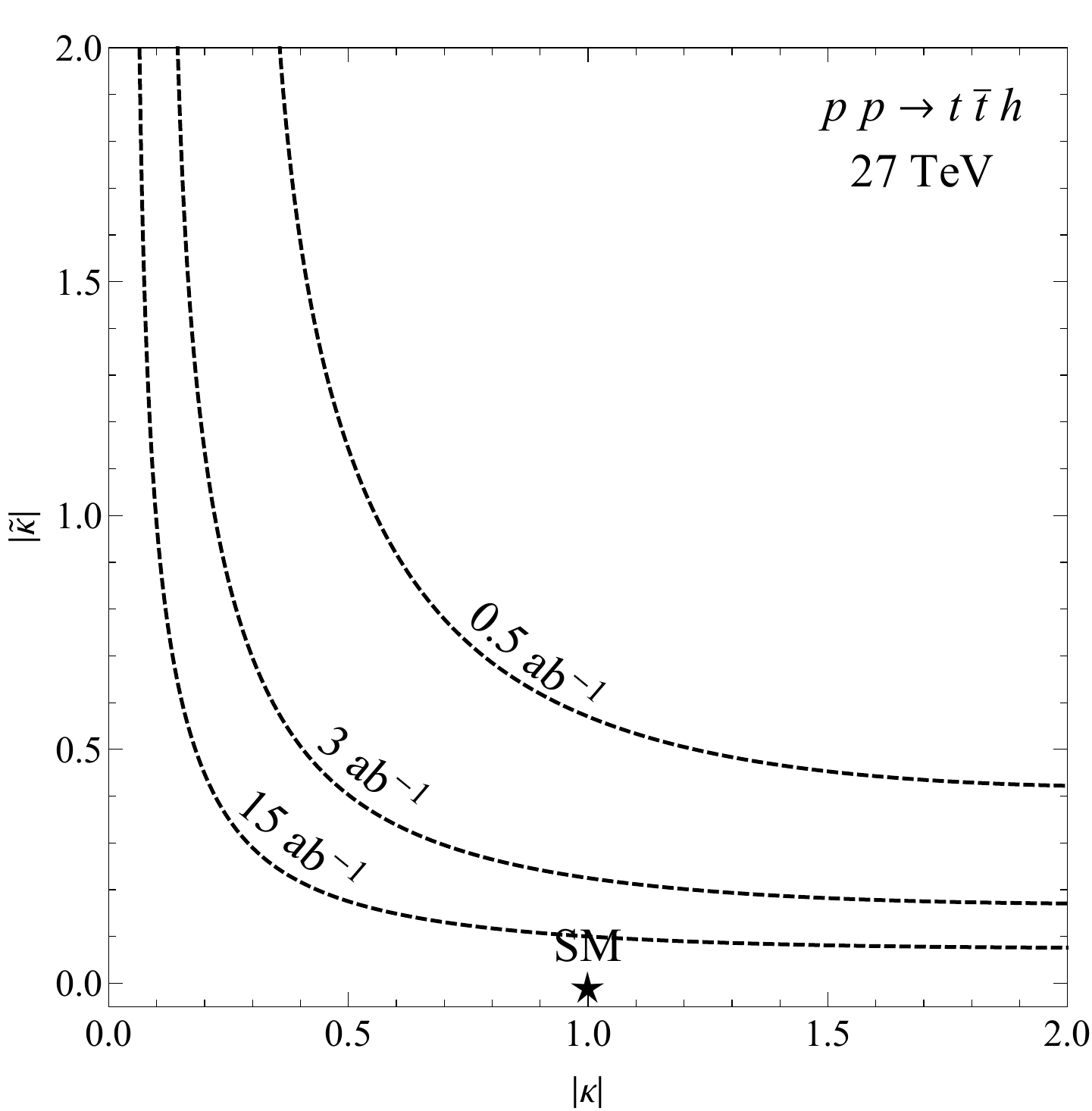}
	\end{tabular}
	\caption{The expected $2\sigma$ exclusion regions for measuring a null result of $\mc{O}_6$ \eqref{eq:Oomega6} are shown for different luminosities at HL-LHC (left) and HE-LHC (right).  }
	
	\label{fig:tthlumbounds}
\end{figure}

\begin{figure}[!t]
	\centering 	
	\includegraphics[width=0.45\linewidth]{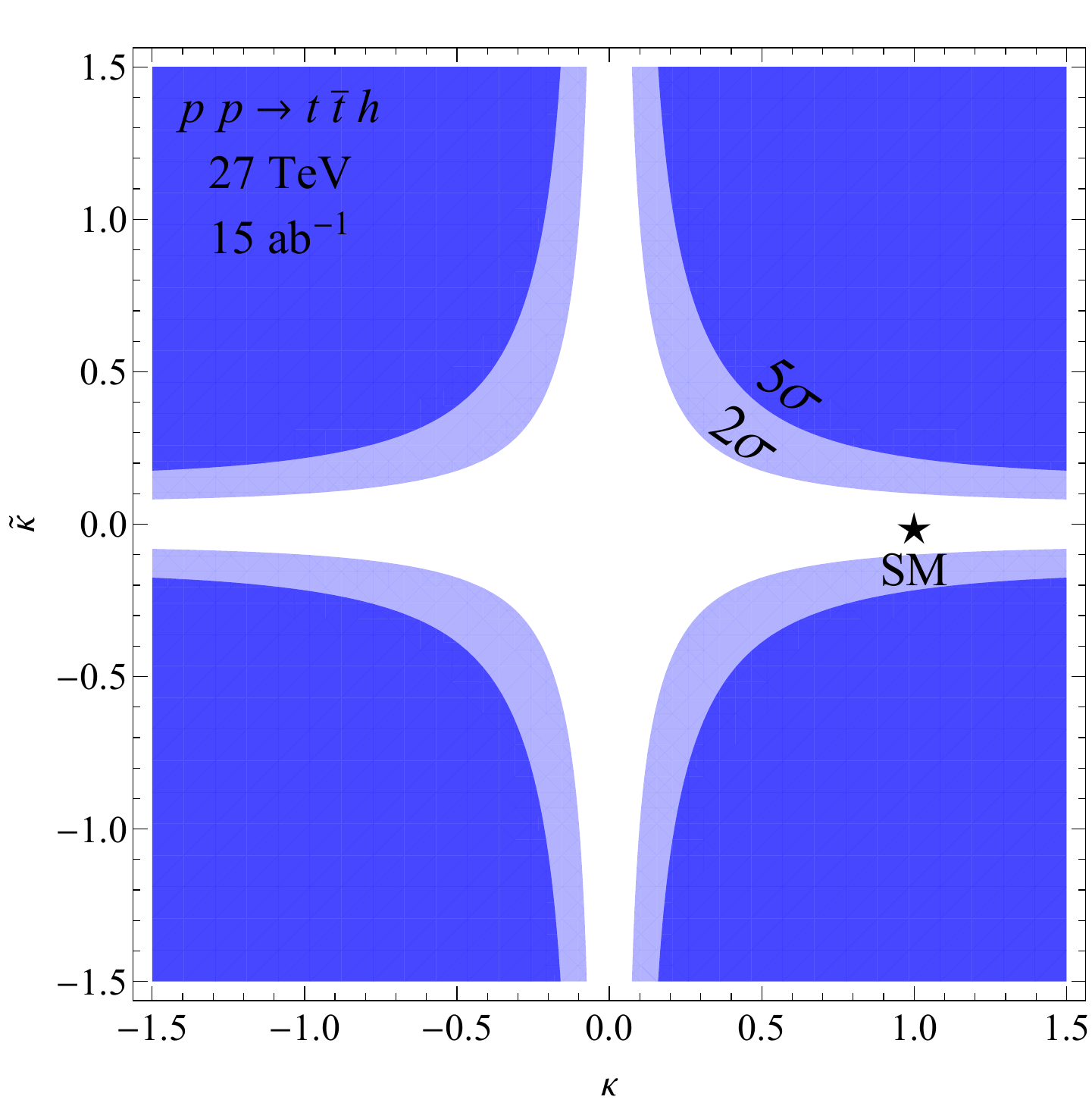}
	\caption{The $2\sigma$ and $5\sigma$ expected exclusion regions for measuring a null result of $\mc{O}_6$ \eqref{eq:Oomega6} at the HE-LHC with $15$ ab$^{-1}$.}	
	\label{fig:tth27bounds}
\end{figure}

\begin{itemize}
	\item 4 or more jets of any flavor with $|\eta(j)|<5$ and $p_T (j) > 20$ GeV.
	\item Of which, at least 3 are $b$-tagged.
	\item Exactly 2 oppositely charged light leptons with $|\eta(\ell)|<2.5$ and $p_T(\ell)>10$~GeV.
\end{itemize}

\noindent Furthermore, in order to identify the $b$-jets from the top-pair decays we count the number of tagged $b$-jets $N_b$ and perform the following selections: if $N_b\ge 4$, we compute the invariant masses $m_{bb}$ of all possible $b$-jet pairs and select the pair with invariant mass closest to the Higgs mass $m_h=125$ GeV. If the selected pair falls inside the Higgs mass window defined by $m_h\pm 15$ GeV we remove the pair from the list of $b$-jets and select from this list the highest $p_T$ $b$-jets as our candidate top quark decay $b$-jets. However if $N_b=3$ we compute all possible invariant masses $m_{bj}$ where $j$ are non-$b$-tagged jets in the event. We select as the $h\to b\bar b$ candidate the $bj$ pair that minimizes $|m_h-m_{bj}|$ and falls inside the Higgs mass window $m_h\pm 15$ GeV. The remaining two $b$-jets are taken as the candidate top quark decay $b$-jets.\\

\begin{figure}[!t]
	\centering 	
	\begin{tabular}{lr}
		\includegraphics[width=0.45\linewidth]{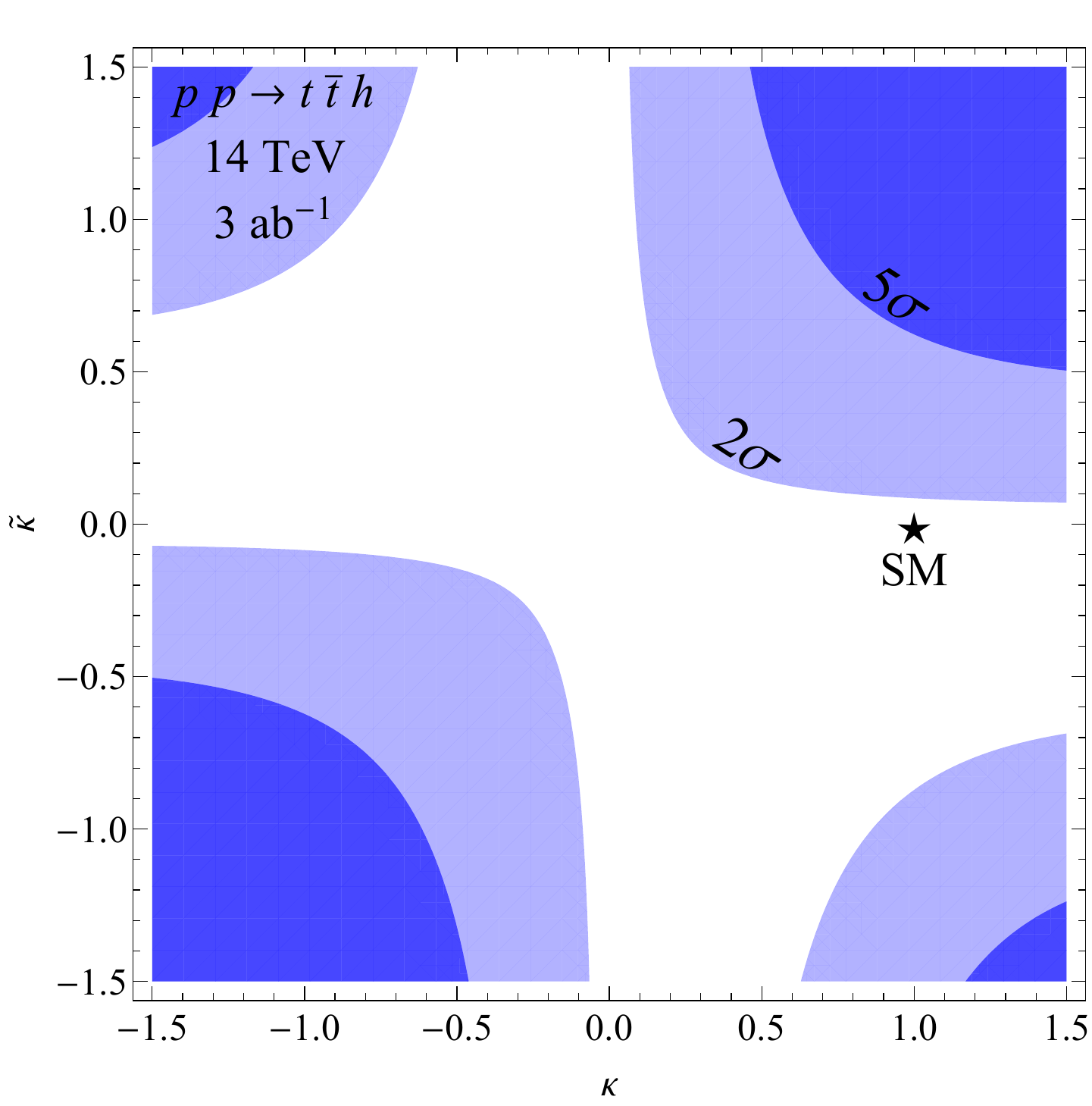} & 
		\includegraphics[width=0.45\linewidth]{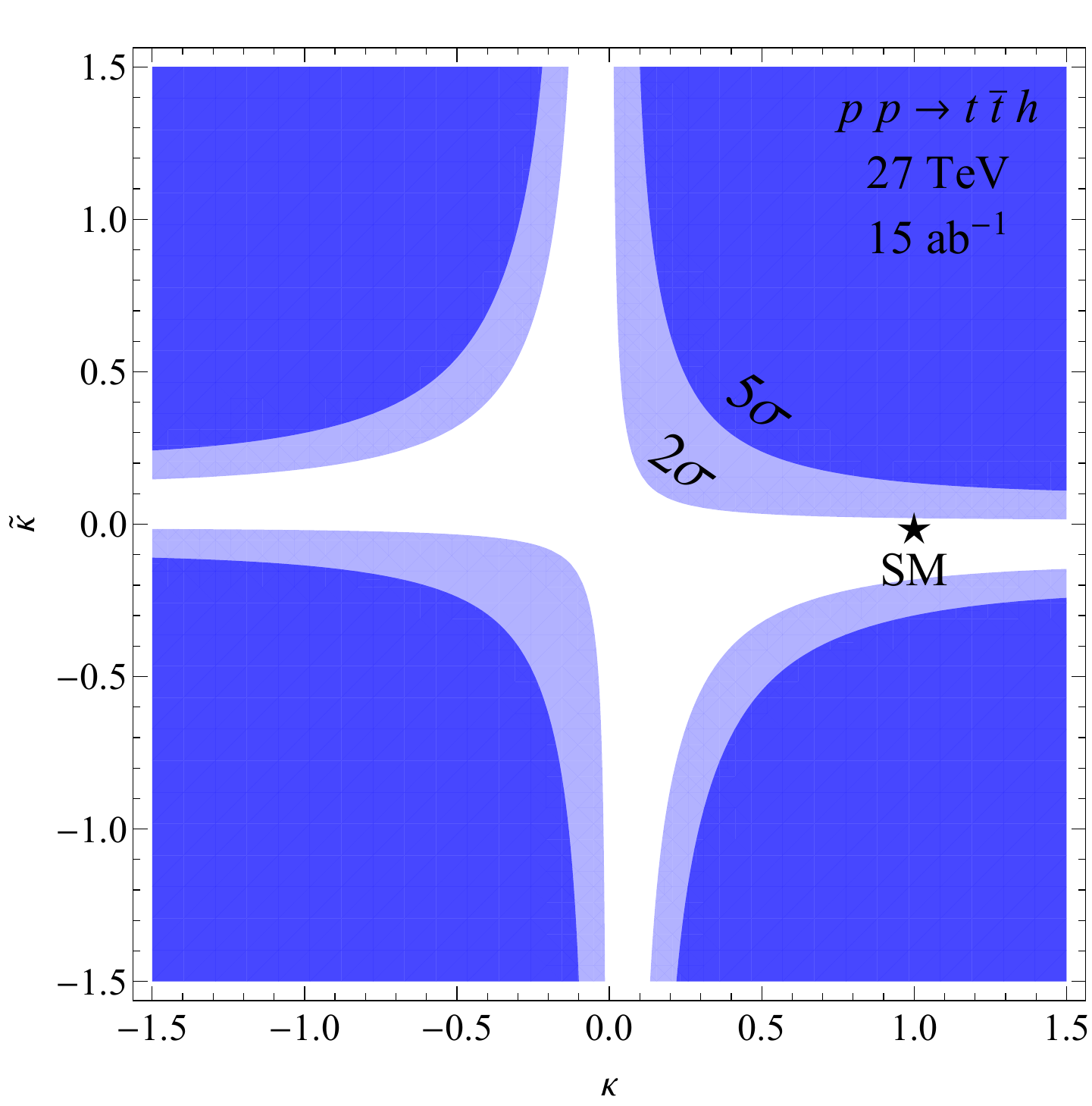}
	\end{tabular}
	\caption{The HL-LHC (3 ab$^{-1}$) and HE-LHC (15 ab$^{-1}$) exclusion regions in the case of measuring a $2\sigma$ positive fluctuation in $\mc{O}_6$ \eqref{eq:Oomega6} (see text for details).}
	\label{fig:tthboundsFluc}
\end{figure}

The reconstruction efficiency of signal events using this approach is $5\%$~$(4.4\%)$ and for background events it is $4.4\%$~$(3.8\%)$ at 14 TeV (27 TeV). We construct a $\chi^2$ for the combined signal and background events. In this case the signal-to-background ratio is much more favorable with $S/\sqrt{B}\sim 32~(128)$ for 3~ab${}^{-1}$ at 14 TeV (15~ab${}^{-1}$ at 27 TeV), so a joint analysis is possible. Results for the $2\sigma$  exclusion regions in the $(\kappa,\tilde\kappa)$ plane are shown in Fig.~\ref{fig:tthlumbounds} for different integrated luminosities at $14$ TeV (left panel) and $27$ TeV (right panel). These results show that the HL-LHC can already probe $\tilde\kappa$ of order 0.5, while the HE-LHC gives an even more promising coverage of parameter space, in particular it is sensitive to $CP$-odd couplings of order $\mc{O}(0.1)$ at high luminosities. In Fig.~\ref{fig:tth27bounds} we provide the $2\sigma$ and $5\sigma$ exclusion regions of the HE-LHC at $15$~ab$^{-1}$. Since the observable $\mc{O}_6$ on the signal behaves as $\propto \kappa\tilde\kappa/(\kappa^2+d \tilde \kappa^2)$ for some constant $d$, the value of $\mc{O}_6$ depends only on $\tilde \kappa/\kappa$. Furthermore, parameter space with small couplings cannot be excluded due to small $S/B$ ratio. These two features lead
to hyperbolic exclusion bounds shown in Fig.~\ref{fig:tth27bounds}. In order to illustrate the sensitivity to the sign of $\tilde\kappa$, we also provide the same exclusion limits in the left panel (right panel) of Fig.~\ref{fig:tthboundsFluc} in the scenario where the measured central value of the observable is $\mc{O}_6=(3.8 \pm 1.9) \times 10^{-4}$ ($\mc{O}_6=(0.8 \pm 0.4) \times 10^{-4}$) at 14 TeV (27 TeV), where the quoted fluctuations and standard deviations are estimated from the statistical error. These measurements, corresponding to a $2\sigma$ excess over the expected null value in the SM.

%
\section{Summary and conclusions}
\label{sec:conclusions}
%
In order to establish, directly and with minimal additional assumptions, the presence of a $CP$-odd component of the top quark Yukawa ($\tilde \kappa$), we have studied manifestly $CP$-odd observables in $th$ and $t\bar t h$ production at the LHC and its prospective upgrades.

For the $t h j$ final states we have relied on the possibility of reconstructing the $t$ quark momentum and accessing the $t$ polarization. We have identified a particular polarization direction which is perpendicular to the $t h$ plane, where the top polarization along this direction would undoubtedly point to the presence of the $CP$-odd coupling $\tilde \kappa$. We have presented a method for optimizing the phase space dependent weight and shown its sensitivity at the HL- and HE-LHC for the semileptonic top and $h \to b\bar b$ mode. The handful of signal events offer discriminating power, sensitive to the sign of $\tilde \kappa$, however the irreducible background due to $t \bar t$+jets severely dilutes the sensitivity of the proposed observable.

On the other hand, $t\bar t h$ production has a considerably larger cross section at LHC energies compared to $t h j$, while suffering more moderately from irreducible backgrounds. Due to the complexity of the final state kinematics with multiple undetected particles we have in this case proposed variables that only depend on the lab-frame accessible momenta and are manifestly $P$- and $CP$-odd. We have identified a single triple product variable that does not rely on $b$-jet charge determination. Finally, among the possible pseudoscalar variables constructed as products of five lab-frame momenta, we have singled out the most sensitive one, $\mc{O}_6$ of Eq.~\eqref{eq:O6}, the sensitivity of which at the HL-LHC with $3\,\mrm{ab}^{-1}$ reaches $\tilde \kappa \sim \mathcal{O}(0.5)$ while the HE-LHC with $15$\,ab${}^{-1}$ would improve this to $\tilde \kappa \sim \mathcal{O}(0.1)$ at $2\sigma$ level. 

Finally, the prospects for directly probing $CP$ violation in the top-quark Yukawa interaction could be potentially further improved by even higher production cross-sections and luminosities offered by the proposed 100\,TeV FCC-hh collider~\cite{Mangano:2016jyj, Contino:2016spe,Benedikt:2018csr}, as well as through better background mitigation techniques, especially in the case of $t h j$ production, and potential phase-space dependent optimization (reweighing) of $CP$-odd observables in $t\bar t h$ production (see e.g. Ref.~\cite{Kraus:2019myc}), all of which we leave for future work.

\begin{acknowledgments}
We would like to thank Jure Zupan for insightful comments. The authors acknowledge support of the Slovenian Research Agency under the core funding grant P1-0035 and J1-8137. A.S. is supported by the Young Researchers Programme of the Slovenian Research Agency under the grant No. 50510, core funding grant P1-0035. D.A.F. has been supported by the Young Researchers Programme of the Slovenian Research Agency under the grant No. 37468, core funding grant P1-0035.  This research was supported by the Munich Institute for Astro- and Particle Physics (MIAPP) which is funded by the Deutsche Forschungsgemeinschaft (DFG, German Research Foundation) under Germany's Excellence Strategy - EXC-2094 - 390783311. We acknowledge support by the COST action CA16201 - “Unraveling new physics at the LHC through the precision frontier”.
\end{acknowledgments}

\appendix

\section{$P$- and $CP$-odd kinematical variables in $pp \to t (\to \ell^+ b \nu) \bar t(\to \ell^- \bar b \bar\nu) h$}
\label{app:tthVars}

Using $\bm{A} = \lm \times \lp$ (see section \ref{sec:labFrameTTH}) we can write down the following
variables:
\begin{equation}
  \label{eq:OmegaL}
  \omega_{\ell}^{\bm{X} \bm{Y}} \equiv \frac{\left[(\lm \times \lp) \cdot \bm{X}\right] \left[(\lm - \lp)\cdot \bm{Y}\right]}{|\lm \times \lp||\bm{X} |\,|\lm-\lp| |\bm{Y}|}, 
\end{equation}
with $\bm{X} = \h, \b+\bbar$ and $\bm{Y} = \h, \lm+\lp, \b+\bbar$,
resulting in 6 possibilities with desired $C$ even and $P$ odd properties:
\begin{align}
    \omega_{\ell}^{h,h} &\sim \left[(\lm \times \lp) \cdot \h\right] \left[(\lm - \lp)\cdot \h\right],\\
    \omega_{\ell}^{h,\ell^- +\ell^+} &\sim \left[(\lm \times \lp) \cdot \h\right] \left[(\lm - \lp)\cdot (\lm+\lp)\right],\\
    \omega_{\ell}^{h,b+\bar b} &\sim \left[(\lm \times \lp) \cdot \h\right] \left[(\lm - \lp)\cdot (\b+\bbar)\right],\\
    \omega_{\ell}^{b+\bar b, h} &\sim \left[(\lm \times \lp) \cdot (\b+\bbar)\right] \left[(\lm - \lp)\cdot \h\right],\\
    \omega_{\ell}^{b+\bar b,\ell^- + \ell^+} &\sim \left[(\lm \times \lp) \cdot (\b+\bbar)\right] \left[(\lm - \lp)\cdot (\lm+\lp)\right],\\
        \omega_6 \equiv \omega_{\ell}^{b+\bar b,b+\bar b} &\sim \left[(\lm \times \lp) \cdot (\b+\bbar)\right] \left[(\lm - \lp)\cdot (\b+\bbar)\right].
\end{align}
Additional possibilities are offered by choosing the
$\bm{A} = \b \times \bbar$ that has to be accompanied by
$\b - \bbar$ (the last column in
Tab.~\ref{tab:CP_transformations_3_momenta}), resulting in variables
\begin{equation}
    \label{eq:OmegaB}
  \omega_{b}^{\bm{X} \bm{Y}} \equiv \frac{\left[(\b \times \bbar) \cdot \bm{X}\right] \left[(\b - \bbar)\cdot \bm{Y}\right]}{|\b \times \bbar||\bm{X}|\,|\b-\bbar||\bm{Y}|}.
\end{equation}
Here $\bm{X} =\h, \lm+\lp$, $\bm{Y} =  \h, \lm+\lp, \b+\bbar$ and the additional combination $\bm{X} = \bm{Y} = \lm-\lp$ makes altogether seven $\omega_b^{\bm{X} \bm{Y}}$ variables:
  \begin{align}
      \omega_{b}^{h,h} &\sim \left[(\b \times \bbar) \cdot \h \right] \left[(\b - \bbar)\cdot \h\right],\\
      \omega_{b}^{h, \ell^- + \ell^+} &\sim \left[(\b \times \bbar) \cdot \h \right] \left[(\b - \bbar)\cdot (\lm+\lp)\right],\\
      \omega_{b}^{h, b+\bar b} &\sim \left[(\b \times \bbar) \cdot \h \right] \left[(\b - \bbar)\cdot (\b+\bbar)\right],\\
      \omega_{b}^{\ell^-+\ell^+, h} &\sim \left[(\b \times \bbar) \cdot (\lm+\lp) \right] \left[(\b - \bbar)\cdot \h\right],\\
      \omega_{b}^{\ell^-+\ell^+,\ell^-+\ell^+} &\sim \left[(\b \times \bbar) \cdot (\lm+\lp) \right] \left[(\b - \bbar)\cdot (\lm+\lp)\right],\\
      \omega_{b}^{\ell^-+\ell^+,b+\bar b} &\sim \left[(\b \times \bbar) \cdot (\lm+\lp) \right] \left[(\b - \bbar)\cdot (\b+\bbar)\right],\\
      \omega_{b}^{\ell^- - \ell^+,\ell^- - \ell^+} &\sim \left[(\b \times \bbar) \cdot (\lm-\lp) \right] \left[(\b - \bbar)\cdot (\lm-\lp)\right].
\end{align}
All $\omega$'s are normalized in a way that links them to the cosines of angles between specific momenta, and implies boundedness, $|\omega| < 1$. In case when $\omega$ is of the form $\bm{A} \cdot \bm{B}\, \bm{A}\cdot\bm{C}$ with $\bm{B} \cdot \bm{C} = 0$ the upper bound is $|\omega| \leq 1/2$.

\bibliography{tYukbib}{}
\end{document}